\documentclass[a4paper,11pt]{article}
\pdfoutput=1 

\usepackage{jheppub} 

\usepackage[T1]{fontenc} 

\usepackage[dvipsnames]{xcolor}
\usepackage{slashed}
\usepackage{amsmath}
\usepackage{comment}

\def\XXint#1#2#3{{\setbox0=\hbox{$#1{#2#3}{\int}$}
     \vcenter{\hbox{$#2#3$}}\kern-.5\wd0}}

\title{\boldmath Hexagons and the classical limit
}

\author{Matheus Fabri and}
\author{Gabriel Lefundes}
\affiliation{Instituto de Física Teórica - UNESP,\\
 Rua Dr. Bento Teobaldo Ferraz 271, 01140-070, São Paulo, Brazil}

\emailAdd{m.fabri@unesp.br}
\emailAdd{g.lefundes@gmail.com}

\abstract{In planar maximally supersymmetric Yang-Mills, we can compute three-point functions at weak coupling using the so-called hexagonalization formalism. The main objects in this framework are called hexagons. We are interested in two sectors of the theory, spanned by operators built entirely from scalar fields, and by spinning operators, respectively. By reviewing the analytic properties of these building blocks, we find new representations for them at weak coupling where the two sectors are on equal footing. We compute the classical limit of the hexagons and of correlation functions in both sectors for some choices of polarizations and our results match previous predictions in the literature.
}

\begin{document} 
\maketitle
\flushbottom


\section{Introduction}

Classical and quantum integrability are paramount to understanding the AdS/CFT correspondence. Over the last two decades, many techniques have been developed to study a plethora of observables in $\mathcal{N}=4$ supersymmetric Yang-Mills (SYM) and its gravity dual at weak, strong, and even at finite coupling \cite{Beisert:2010jr,Bombardelli:2016rwb}. One example of outstanding success is the computation of the spectrum of single-trace operators to all-loops in perturbation theory, which revealed rich integrable structures as the quantum spectral curve \cite{Gromov:2013pga}.

\begin{figure}[t]
\centering
\includegraphics[width=.75\textwidth]{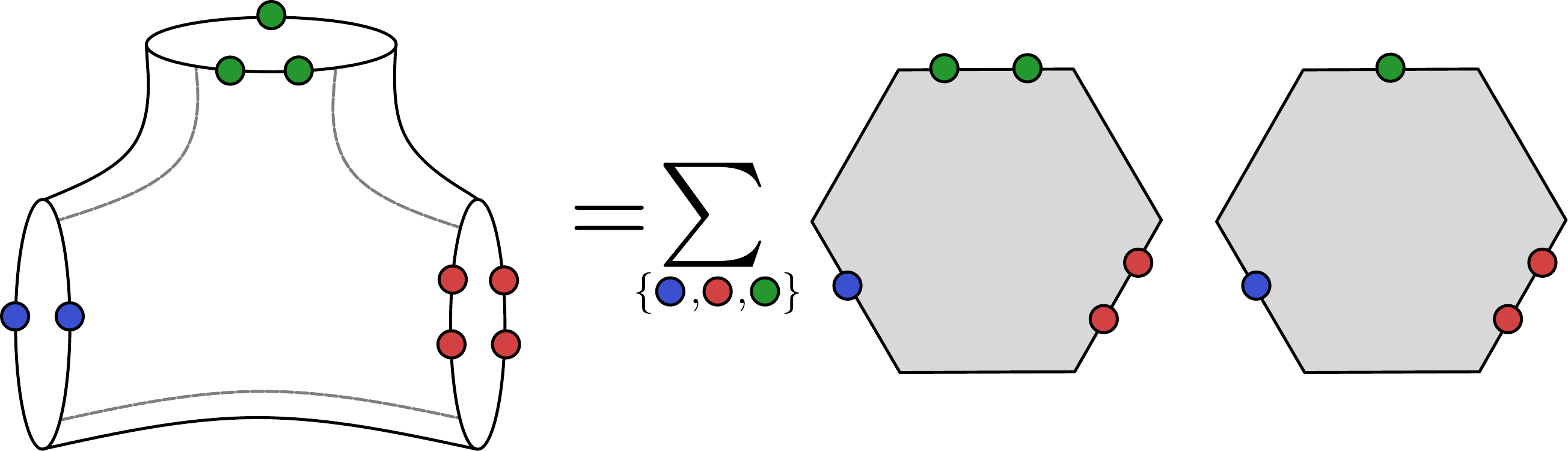}
\caption{\label{fig:hexagons}  Hexagonalization. On the left-hand side we have a structure constant represented by a pair of pants, and its excitations, represented by the colored dots. To compute its value, we sum over all possible ways of splitting the excitations in the front and back hexagons.
}
\end{figure}

These developments build on the remarkable observation that we can recast the one-loop dilatation operator of $\mathcal{N}=4$ SYM as the Hamiltonian of an integrable spin-chain. There, we parametrize single-trace operators by the eigenvalues of the system \cite{Minahan:2002ve}. In this picture, the so-called tailoring procedure emerged, allowing one to find the three-point functions of single-trace operators at weak coupling in terms of the inner products of spin-chain states, computed using quantum integrability \cite{Escobedo:2010xs,Escobedo:2011xw,Gromov:2011jh,Gromov:2012uv,Vieira:2013wya,Kazama:2014sxa}. More recently, it was conjectured that one could compute these structure constants by decomposing them into building blocks called hexagons (see Figure \ref{fig:hexagons}), which are fixed by the assumed all-loop integrability \cite{Basso:2015zoa}.

Although the tailoring formalism and hexagonalization compute the same structure constants in the SU(2) sector, their equivalence was not proven. In this work we amend this by proving, in full generality, that both are indeed equivalent. Also by working with the analytic properties of the hexagons we find new representations for them at tree level in both the SU(2) and SL(2) sectors. This could hint to a new representation of them at finite coupling. Also these new expressions are computationally less costly, since we do not need to do multiple tensor contractions to compute them.

In this work we also consider the classical limit which is central at strong coupling. Indeed, in this regime, the three-point functions are computed in a very different fashion: we use AdS/CFT to map them to the scattering of closed strings in AdS. The non-linear sigma model on the gravity dual is classically integrable, and, for strings with large charges, the correlation functions can be evaluated by saddle-point; these `heavy' strings correspond in $\mathcal{N} = 4$ SYM to operators with large charges \cite{Tseytlin:2010jv}. 

\begin{figure}[t]
\centering
\includegraphics[width=.5\textwidth]{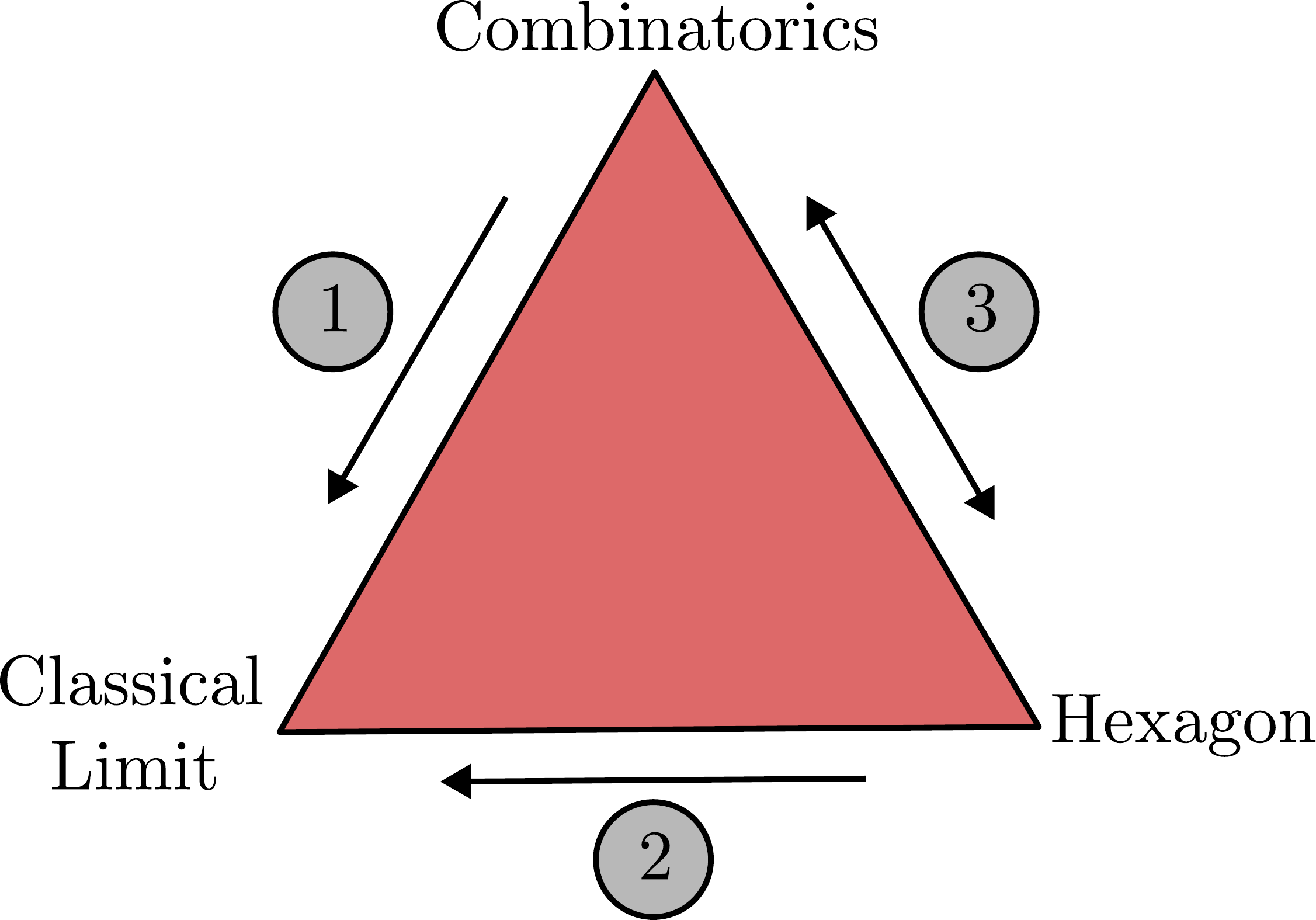}
\caption{\label{fig:triangle} Triangle of relations. In this paper we link different formalisms and limits. From tailoring (Combinatorics) we compute the classical limit of the so-called type I-I-I correlators (1) and establish the relation with hexagonalization (3). Also we compute the classical limit of SL(2) and SU(2) correlators from hexagonalization (2) using new hexagon representations.
}
\end{figure}

However the classical limit also plays a role at weak coupling. One instance of this is the famous Frolov-Tseytlin limit, where at strong coupling and asymptotically large charges we find the same results at weak coupling under the expansion in a appropriate parameter\footnote{However there are some observables where this duality breaks at some loop order, see for example \cite{Minahan:2005qj}.} \cite{Frolov:2003xy}. Another instance where we can connect these a priori unrelated appearances of integrable structures is by taking the classical limit of the spin-chain associated with heavy single-trace operators. In this limit, we have a large number of Bethe roots that condense into macroscopic branch cuts in the complex plane, resembling the finite-gap solutions of the classical string sigma model \cite{Kazakov:2004qf,Kazakov:2004nh}. In the particular case of operators in a SU(2) sector of $\mathcal{N}=4$ SYM, the collective dynamics of excitations can be described macroscopically by the Landau-Lifschitz model, which makes the relationship with the strong coupling framework very explicit \cite{Kazama:2016cfl}. In this work we compute the classical limit of a special class of correlators in the SU(2) sector: the type I-I-I correlators. Our derivation is done directly from the microscopic formulas using recursion relations we derived for the hexagons analogous to the ones in \cite{Bercini:2022gvs}.  We use the same methods to also compute the classical limit of SL(2) three-point functions for some fixed polarizations of the external operators. The results of this paper are summarized in Figure \ref{fig:triangle}.

This work is divided in the following way. In Sec. \ref{sec:sec2} we review some results that we will use throughout the paper. Then in Sec. \ref{sec:new-hex-rep} we describe the new hexagon representations in the SU(2) and SL(2) sector. The classical limit is then discussed in Sec. \ref{sec:sec3} for both sectors and in Sec. \ref{sec:sec5} we end up with some concluding remarks and future prospects. We relegate some technical details to the appendices. 

\section{Setup and review}

\label{sec:sec2}

In this section we review the tailoring procedure for computing 
three-point functions for operators with polarizations lying in the SO(4) sector. Then we review the spinning hexagon method to compute structure constants in the SL(2) sector and we close with the classical limit of spin-chains and present some known results in the literature in this direction.


\subsection{Double spin-chain formalism}

The double spin-chain approach introduced in \cite{Kazama:2014sxa} to compute structure constants in the SO(4) sector consists basically of mapping the operators to entangled spin-chain states, cutting them, and contracting the Bethe states. In this generalization of the tailoring formalism, the authors in  \cite{Kazama:2014sxa} define a double spin-chain using the decomposition of the fundamental SO(4) representation in terms of a tensor product of two doublet SU(2) representations. This idea allowed for the description of correlators with operators built on top of distinct vacua, which is necessary to study more general configurations in the SO(4) sector.

We start by mapping each of the scalars $\{Z,Y,\bar{Z},\bar{Y}\}$ in the SO(4) sector to a double spin-chain state:
\begin{equation}
\begin{array}{cc}
 Z \rightarrow |\uparrow\rangle\otimes|\uparrow\rangle, &  \Bar{Z} \rightarrow |\downarrow\rangle\otimes|\downarrow\rangle,  \\
Y  \rightarrow |\uparrow\rangle\otimes|\downarrow\rangle , &  \Bar{Y} \rightarrow |\downarrow\rangle\otimes|\uparrow\rangle.
\end{array}
\end{equation}
In field theory language, one treats $Z$ and $\bar{Z}$ as vacua and $Y$ and $\bar{Y}$ as excitations. To build the excited states, we put excitations on the left or right spin-chain. Note that we can not excite both sectors at once since this would give an operator outside the SO(4) sector. If we excite the left spin-chain, we have a \textit{type I} operator; otherwise, we have a \textit{type II} operator. Finally, we rotate the vacua with a SU(2)$_L \times $SU(2)$_R$ transformation to consider different polarizations, parametrized by complex numbers $z_i$ and $\tilde{z}_i$.

Instead of explaining the whole cutting and tailoring procedure of  \cite{Kazama:2014sxa} we show here the full result and then explain each piece of it. The structure constants of three non-BPS operators in the SO(4) sector of $\mathcal{N}=4$ SYM at weak coupling can be written as 
\begin{equation}
\label{structure}
    C^{\bullet\bullet\bullet} = \frac{\sqrt{\ell_1 \ell_2 \ell_3}}{N_c \ \mathcal{N}_1 \mathcal{N}_2 \mathcal{N}_3} c_L c_R,
\end{equation}
where $\mathcal{N}_j$ is the normalization of each operator that is given by the Gaudin norm (see Appendix \ref{AppA}), $\ell_j$ is the length of each operator, and $N_c$ is the number of colors in the gauge group.  Here $c_L$ and $c_R$ denote the left and right spin-chain contributions, respectively. Let $\mathbf{u}_k$ be the set of roots for each operator. Then, the left chain factor is given by
\begin{equation}
\label{scdef}
    c_L = \prod\limits_{i=1}^{3}\left(\frac{1}{1+|z_i |^2}\right)^{\ell_i /2 - M_i} \sum\limits_{\alpha_{k}\cup\Bar{\alpha}_{k} = \mathbf{u}_{k}} z_{12}^{\ell_{12}-|\alpha_{1}|-|\Bar{\alpha}_{2}|}\ z_{13}^{\ell_{13}-|\alpha_{1}|-|\Bar{\alpha}_{3}|} \ z_{23}^{\ell_{23}-|\alpha_{2}|-|\Bar{\alpha}_{3}|}\ \mathcal{D}(\alpha_j, \bar{\alpha}_j),
\end{equation}
where
\begin{align}
\label{Ddef}
    &\mathcal{D}(\alpha_j, \bar{\alpha}_j) =  (-1)^{|\alpha_{1}|+|\alpha_{2}|+|\alpha_{3}|}\times \prod\limits_{i=1}^{3} \textcolor{red}{H_{\ell_i} (\alpha_{i} , \Bar{\alpha}_{i})} \\
    & \hspace{4cm}  \times  \textcolor{blue}{Z_p (\alpha_{1} \cup \Bar{\alpha}_{3} |\ell_{13}) Z_p (\alpha_{2} \cup \Bar{\alpha}_{1} |\ell_{12})} \textcolor{blue}{ Z_p (\alpha_{3} \cup \Bar{\alpha}_{2} |\ell_{23})} , \nonumber \\
    & \textcolor{red}{H_{\ell} (\alpha_l , \alpha_r)} =  
     \prod\limits_{u\in\alpha_l}\prod\limits_{v\in\alpha_r} \frac{u-v+i}{u-v} \left( u- \frac{i}{2} \right)^{\ell_r} \left( v+ \frac{i}{2} \right)^{\ell_l}, \\
    & \textcolor{blue}{Z_p (\mathbf{u}|\ell)} =  \frac{\prod\limits_{j} (u_j - i/2)^{\ell}}{\prod\limits_{i<j} (u_i - u_j)} \times \det \left(u_b^{a-1} -\left(\frac{u_b + i/2}{u_b - i/2}\right)^{\ell} (u_{b}-i)^{a-1} \right).
\end{align}
Here, $z_{ij} = z_{i} - z_{j}$, $\ell_{ij}$ is the bridge length between operators $i$ and $j$, $|\alpha|$ denotes the length of the set $\alpha$, and $M_i$ is the number of excitations in the $i$-th operator.  The terms in \textcolor{red}{red} come from cutting the states of each chain, while the terms in \textcolor{blue}{blue} come from the overlap between the left and right subchains of adjacent operators\footnote{The overlaps can be written in terms of the partial domain wall partition functions \cite{Kazama:2014sxa,Foda:2012yg,Foda:2011rr}, which are the usual domain wall partition functions when some of the rapidities are taken to infinity.}. Finally, we sum over all splittings of the Bethe roots. For $c_R$ the definition is the same with the corresponding $\Tilde{z}_{ij}$ variables.

\begin{figure}[t]
\centering
\includegraphics[width=.9\textwidth]{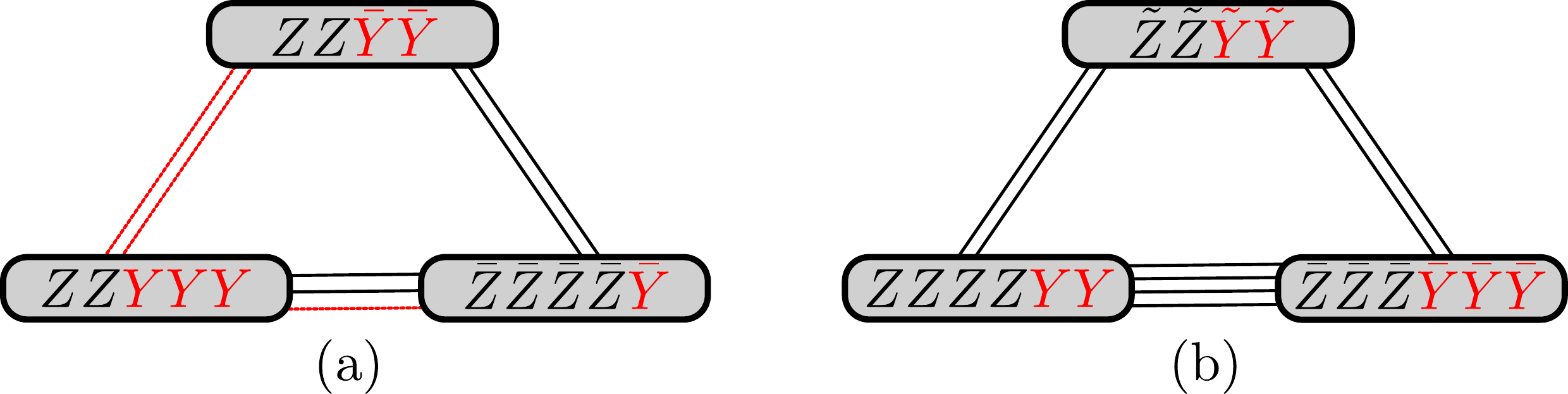}
\caption{\label{fig:examples} Examples of type I-I-II and I-I-I correlators. In (a) we have an example of a type I-I-II correlator analyzed in \cite{Gromov:2011jh}. Now in (b) we have a type I-I-I correlator where $\tilde{Z}=Z+\bar{Z}+Y-\bar{Y}$ and $\tilde{Y}=Y+\bar{Z}$. This structure constant was analyzed in \cite{Basso:2015zoa}.}
\end{figure}

We can separate the correlators into two classes: if all operators are of the same type we have a \textit{type I-I-I} or \textit{pure} correlator; otherwise, we have a \textit{mixed} or \textit{type I-I-II} correlation function. Examples of these correlators are given in Figure \ref{fig:examples}. If the operators are on-shell, i.e., they satisfy the Bethe equations, the $z_{ij}$ dependence is completely fixed by symmetry \cite{Kazama:2014sxa}. Then, dropping the $L$ label, we have
\begin{equation}
\label{su2inv}
    c =  \prod\limits_{i=1}^{3}\left(\frac{1}{1+|z_i |^2}\right)^{\ell_i /2 - M_i}  z_{12}^{\ell_{12}-M_{12}} z_{23}^{\ell_{23}-M_{23}} z_{31}^{\ell_{13}-M_{13}}\ \mathcal{G},
\end{equation}
where $M_{ij} = M_{i} + M_{j}  - M_{k}$, and $\mathcal{G}$ is independent of $z_{ij}$. Using the Bethe equations, we were able to write an expression for $\mathcal{G}$ which is independent of $z_{i j}$ even for off-shell operators \cite{Basso:2015zoa}. This rewrite will allow us later to take the classical limit of the pure correlators in Section \ref{sec:pure-clas-lim}. Then, we can write 

\begin{equation}
\mathcal{G} = \sum_{\alpha_k \cup \Bar{\alpha}_k = \mathbf{u}_{k}} z_{1 2}^{M_{12} - |\alpha_2|- |\bar{\alpha}_1| } z_{2 3}^{M_{23} - |\alpha_3|- |\bar{\alpha}_2| } z_{3 1}^{M_{31} - |\alpha_1|- |\bar{\alpha}_3| } \ \tilde{\mathcal{D}}(\alpha_j, \bar{\alpha}_j),
\label{eq:fullsc}
\end{equation}
where 
\begin{multline}
\label{defDinv}
    \tilde{\mathcal{D}}(\alpha_j, \bar{\alpha}_j)  = (-1)^{|\bar{\alpha}_1| + |\bar{\alpha}_2| + |\bar{\alpha}_3|} \sum_{\substack{\beta_1 \cup \bar{\beta}_1 = \bar{\alpha}_1 \cup \alpha_2 \\ \beta_2 \cup \bar{\beta}_2 = \bar{\alpha}_2 \cup \alpha_3 \\ \beta_3 \cup \bar{\beta}_3 = \bar{\alpha}_3 \cup \alpha_1}}(-1)^{|\beta_1| + |\beta_2| + |\beta_3|}   \\
    \times a_{\ell_{12}}(\bar{\beta}_1 \cap \alpha_2) a_{\ell_{23}}(\bar{\beta}_2 \cap \alpha_3)
    a_{\ell_{31}}(\bar{\beta}_3 \cap \alpha_1) a_{\ell_{31}}(\beta_1 \cap \bar{\alpha}_1) a_{\ell_{12}}(\beta_2 \cap \bar{\alpha}_2) a_{\ell_{23}}(\beta_3 \cap \bar{\alpha}_3) \\ \times S\left(\mathbf{u}_{1},\beta_1 \cap\bar{\alpha}_1\right) S\left(\mathbf{u}_{2},\beta_2 \cap\bar{\alpha}_2\right)
    S\left(\mathbf{u}_{3},\beta_3 \cap\bar{\alpha}_3\right)
    \\ \times f(\bar{\alpha}_1,\alpha_1)f(\bar{\alpha}_2,\alpha_2) f(\bar{\alpha}_3,\alpha_3) f(\beta_1,\bar{\beta}_1) f(\beta_2,\bar{\beta}_2) f(\beta_3,\bar{\beta}_3),
\end{multline}
and 
\begin{align}
    & f(\mathbf{x},\mathbf{y}) = \prod_{u\in \mathbf{x}} \prod_{v\in \mathbf{y}} \left(\frac{u-v+i}{u-v} \right), \\
    & S(\mathbf{x},\mathbf{y}) =\frac{f(\mathbf{x},\mathbf{y})}{f(\mathbf{y},\mathbf{x})} = \prod_{u\in \mathbf{x}} \prod_{v\in \mathbf{y}} \left(\frac{u-v+i}{u-v-i} \right), \\
    & a_{\ell}(\mathbf{x}) = \prod_{u\in \mathbf{x}} \left(\frac{u+i/2}{u-i/2} \right)^{\ell}.
\end{align}
The derivation of these formulas is done in details in Appendix \ref{AppB}. Physically, $S(u,v)$ is the SU(2) spin-chain S-matrix, $a_\ell$ is $e^{iP \ell}$, where $P$ is the total momentum of the excitations, and $f(u,v)$ is the inverse of the hexagon scalar factor at tree level. The reader can wonder if the independence on the polarizations is true since they still appear in \eqref{eq:fullsc}, however we checked in many examples that this property indeed holds for generic roots and not only on-shell ones.

For three-point functions of type I-I-II, the fact that $\mathcal{G}$ does not depend on the polarizations allow us to reduce the sum \eqref{eq:fullsc} to a single term and reproduce the factorized expressions in \cite{Escobedo:2011xw,Kostov:2012jr} which was done in \cite{Kazama:2014sxa}. However, for type I-I-I, the sums are coupled, which makes it very hard to compute their classical limit. In this work, we were able to find factorized expressions for type I-I-I when the roots of different operators are macroscopically distant.

\subsection{Spinning hexagons}
\label{sec:spinning-hex}

Since we already detailed the compact sector, here we focus on three-point functions of three non-BPS single-trace operators in the SL(2) sector of the theory. An operator with spin $J$ in this sector is given by
\begin{equation}
    \mathcal{O}_J (x,L,R) =  L_{\alpha_1} R_{\dot{\alpha}_1}\cdots L_{\alpha_J} R_{\dot{\alpha}_J} \ \mathcal{O}_{J}^{\alpha_1 \dot{\alpha}_1 \cdots \alpha_J \dot{\alpha}_J} (x),
\end{equation}
where $\mathcal{O}_{J}^{\alpha_1 \dot{\alpha}_1 \cdots \alpha_S \dot{\alpha}_S} (x)$ is a BPS vacuum of length $\ell$ with $J$ covariant derivatives $\mathcal{D}^{\alpha \dot{\alpha}}$ applied on it. Each operator is characterized by the polarization spinors $L_{\alpha}$ and  $R_{\dot{\alpha}}$, from which we can construct the SL(2) invariants 
\begin{equation*}
    H_{i j} = \langle L_i, R_j \rangle \langle L_j, R_i \rangle, \ \ \ V_i = \langle L_i, R_i \rangle \ \ \ \textrm{and} \ \ \ t_{i j} = \frac{H_{i j}}{V_i V_j}.
\end{equation*}
Due to SL(2) invariance the structure constant can depend only on these and furthermore after some redefinitions it is a polynomial in the $t_{i j}$ variables \cite{Bercini:2020msp}.

\begin{figure}[t]
\centering
\includegraphics[width=.65\textwidth]{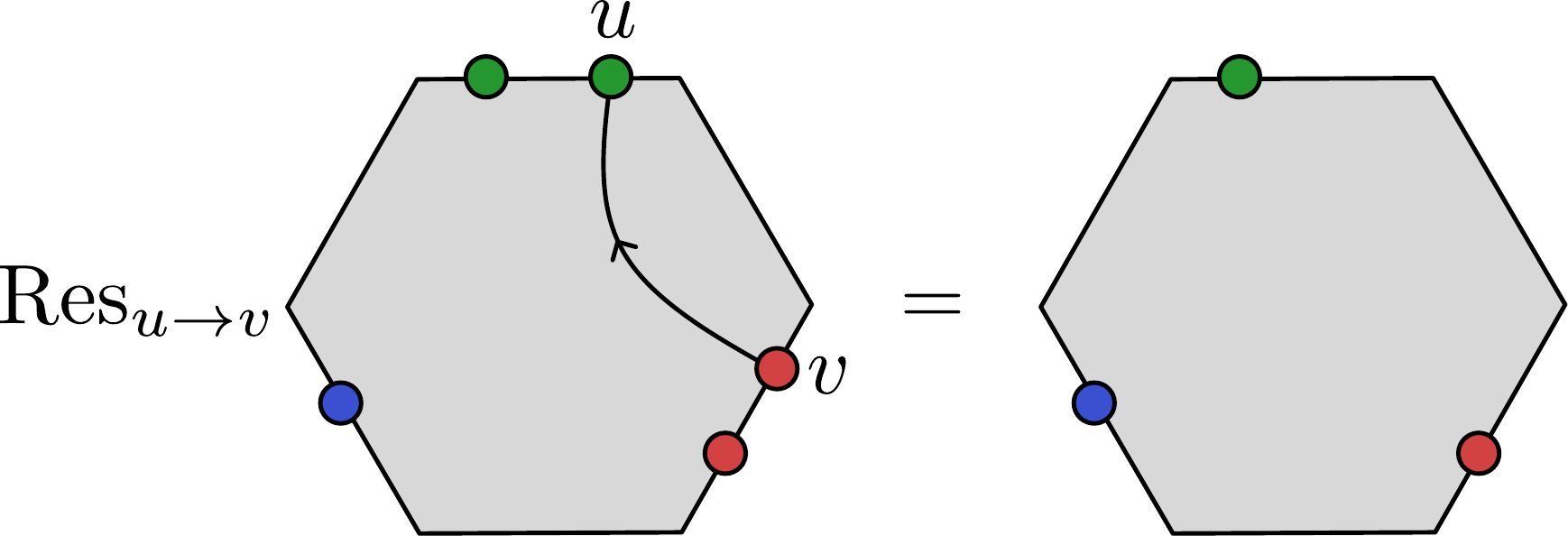}
\caption{\label{fig:decoupling} Decoupling axiom. When two roots of distinct operators are equal we have a pole whose residue is a hexagon without these two excitations.}
\end{figure}

The correlators in this sector can be computed from hexagonalization at finite coupling. An improvement in this direction was given in \cite{Bercini:2022gvs}. Here the building blocks of the hexagonalization approach, the so-called hexagons, were reinterpreted as partition functions in a Kagome lattice. Although these still hard to compute at finite coupling, at tree level we can find them using recursion relations.  These stem from the hexagon decoupling axioms, here shown in Figure \ref{fig:decoupling}. The recursion relations can be written as

\begin{align}
    &\text{Res}_{y \rightarrow x} \ \mathcal{H}(\mathbf{u}_1 \cup \{x\},\mathbf{u}_2 \cup \{y\},\mathbf{u}_3) = - i t_{1 2}\ f(x,\mathbf{u}_1) f(\mathbf{u}_2,x) \mathcal{H}(\mathbf{u}_1,\mathbf{u}_2,\mathbf{u}_3), \label{eq:eq1-rr} \\
    &\text{Res}_{y \rightarrow x} \ \mathcal{H}(\mathbf{u}_1,\mathbf{u}_2 \cup \{x\},\mathbf{u}_3\cup \{y\}) = - i t_{2 3}\ f(x,\mathbf{u}_2) f(\mathbf{u}_3,x) \mathcal{H}(\mathbf{u}_1,\mathbf{u}_2,\mathbf{u}_3), \label{eq:eq2-rr}\\
    &\text{Res}_{y \rightarrow x} \ \mathcal{H}(\mathbf{u}_1 \cup \{y\},\mathbf{u}_2,\mathbf{u}_3 \cup \{x\}) = - i t_{3 1}\ f(x,\mathbf{u}_3) f(\mathbf{u}_1,x) \mathcal{H}(\mathbf{u}_1,\mathbf{u}_2,\mathbf{u}_3),\label{eq:eq3-rr}
\end{align}
where $\mathcal{H}(\mathbf{u}_1,\mathbf{u}_2,\mathbf{u}_3)$ is the hexagon form factor for these three root sets and here for SL(2)
\begin{align}
   f(\mathbf{x},\mathbf{y}) = \prod_{u\in \mathbf{x}} \prod_{v\in \mathbf{y}} \left( \frac{u-v-i}{u-v} \right).  
\end{align}
Note the sign change with respect to the SU(2) definition. This is due to the fact that the SL(2) and SU(2) S-matrices are inverse of each other. The remaining set of decoupling relations are
\begin{align}
    \text{lim}_{x \rightarrow \infty} \mathcal{H}(\mathbf{u}_1 \cup \{x\},\mathbf{u}_2,\mathbf{u}_3) = \mathcal{H}(\mathbf{u}_1,\mathbf{u}_2,\mathbf{u}_3), \label{eq:dec1} \\
    \text{lim}_{x \rightarrow \infty} \mathcal{H}(\mathbf{u}_1,\mathbf{u}_2\cup \{x\},\mathbf{u}_3) = \mathcal{H}(\mathbf{u}_1,\mathbf{u}_2,\mathbf{u}_3), \label{eq:dec2} \\
    \text{lim}_{x \rightarrow \infty} \mathcal{H}(\mathbf{u}_1 ,\mathbf{u}_2,\mathbf{u}_3\cup \{x\}) = \mathcal{H}(\mathbf{u}_1,\mathbf{u}_2,\mathbf{u}_3).
    \label{eq:dec3}
\end{align}
Turns out that at tree level the hexagon form factor is a rational function of the rapidities and it can be totally fixed by its residues and asymptotic behavior, therefore using \eqref{eq:eq1-rr} - \eqref{eq:eq3-rr} and \eqref{eq:dec1} - \eqref{eq:dec3} it can be completely specified. Our recursion relations differ somewhat from the ones discovered in \cite{Bercini:2022gvs}, due to the fact that we redefine the hexagon to obtain simpler recursion relations and match notation with the SU(2) case, this redefinition is detailed in Appendix \ref{App:redefinitions}. In Section \ref{sec:su2-newrep} we will derive similar relations for the SU(2) case.

After finding $\mathcal{H}$, we can recover the full three-point function through
\begin{multline}
    \label{eq:Gspinning}
    \mathcal{G}(\mathbf{u}_1,\mathbf{u}_2,\mathbf{u}_3) = \sum_{\delta_i \cup \bar{\delta}_i = \mathbf{u}_i} (-1)^{|\delta_1| + |\delta_2| + |\delta_3|} f(\bar{\delta}_1,\delta_1) f(\bar{\delta}_2,\delta_2) f(\bar{\delta}_3,\delta_3) \\ \times  a_{\ell_{31}}(\delta_1) a_{\ell_{12}}(\delta_2) a_{\ell_{23}}(\delta_3)  \mathcal{H}(\delta_1,\delta_3,\delta_2) \mathcal{H}(\bar{\delta}_1,\bar{\delta}_2,\bar{\delta}_3).
\end{multline}
Note here that unlike the SU(2) case, the polarization tensor structure is not fixed, \textit{id est}, in \eqref{eq:Gspinning} there are many polynomial terms in $t_{ij}$ and not a single one as in the SU(2) case \cite{Costa:2011mg}. In Section \ref{sec:sl2-newrep} we will derive a closed form expression for the hexagons that satisfy all these recursion relations.

\subsection{Classical limit}

An important regime to analyze physical quantities in AdS/CFT is the so-called classical limit. In this case one consider asymptotically heavy operators in $\mathcal{N}=4$ SYM with infinitely many excitations on each operator. More specifically, the classical limit in the SU(2) sector is defined as 
\begin{equation*}
    \ell \sim M \sim u_i \rightarrow\infty.
\end{equation*}
In this regime, the roots of each operator grows with $\ell$ and with separation of order $\mathcal{O}(1)$. Thus after a $1/\ell$ reescaling the roots' separation are of order $\mathcal{O}(1/\ell)$ and the set of roots condense into a continuous contour $\mathcal{C}_i$ \cite{Kazakov:2004qf}. This root condensation is shown in Figure \ref{fig:class-limit}. Then each root set is characterized by a density $\rho_i(u)$ in the complex plane defined at the contour $\mathcal{C}_i$. Using these facts we can construct the classical limit of $f$ and $a_\ell$ \cite{Gromov:2011jh}:

\begin{figure}[t]
\centering
\includegraphics[scale=0.3]{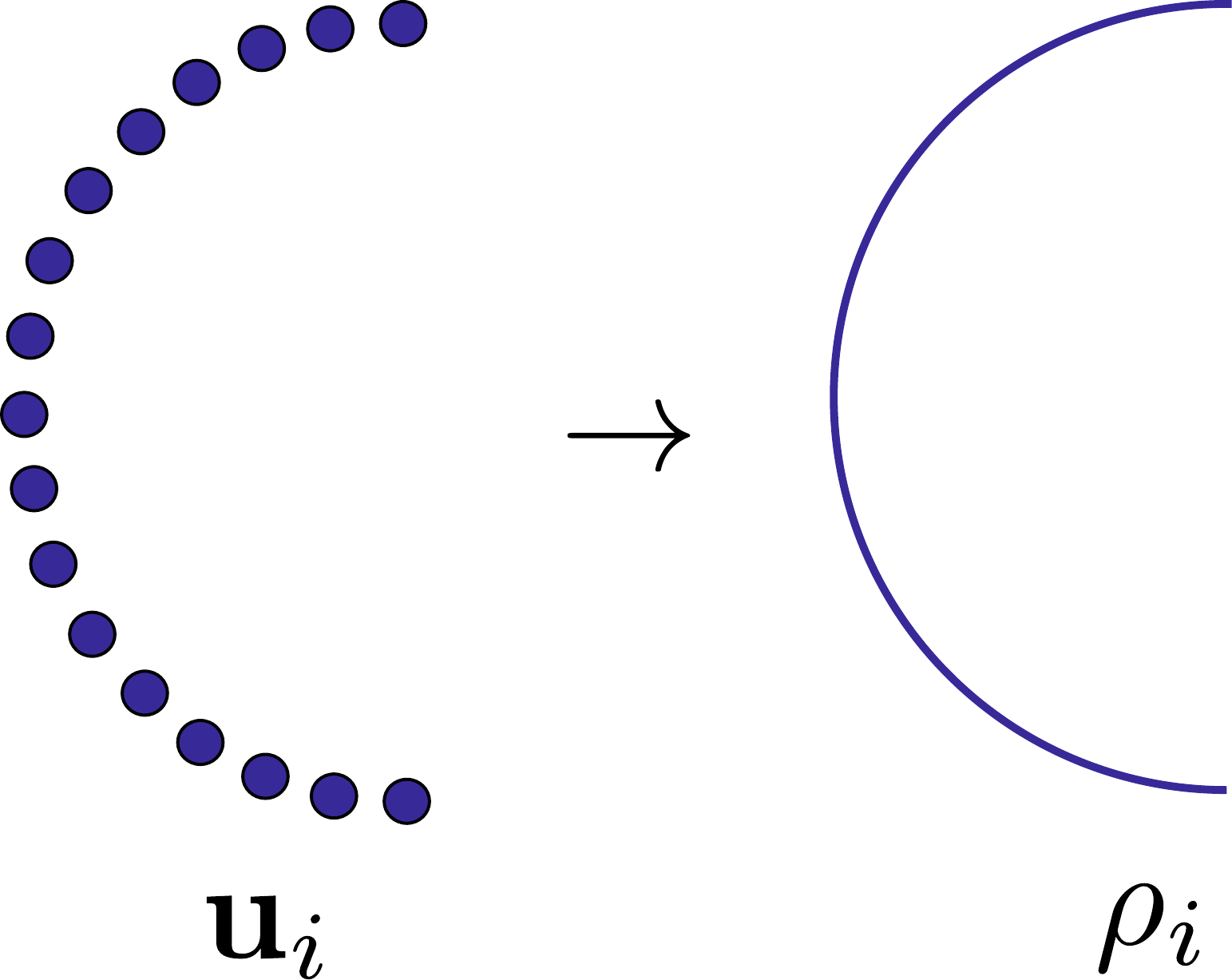}
\caption{\label{fig:class-limit} Condensation of roots. In the classical limit, the roots condense into cuts and we describe the set of roots $\mathbf{u}_i$ by a root density $\rho_i$ in the complex plane.}
\end{figure}

\begin{align}
    f(z,\mathbf{u}) \rightarrow e^{i G(z)} \ \ \ \  \text{and} \ \ \ \ a_{\ell}(z) \rightarrow e^{i \frac{\ell}{z}},
\end{align}
where $G(z)$ is the resolvent
\begin{align}
    G(z) = \int_{\mathcal{C}}  dx\ \frac{\rho (x)}{z-x}.
\end{align}
Unlike the density, it is clear that the resolvent is defined at the entire complex plane modulo the root contours. The Bethe equations in the classical limit imply that the resolvent is discontinuous in $\mathcal{C}_i$, due to this we also call $\mathcal{C}_i$ as the roots' cut \cite{Kazakov:2004qf}. Remarkably, the structure constants depend not on the densities explicitly, but only through the quasimomentum
\begin{align}
    p(z) = G(z) - \frac{\ell}{2 z}.
\end{align}
Note that the quasimomentum is a meromorphic function with singularities at the origin and branch cuts at $\mathcal{C}$. 

Now we present some known results in the literature. The classical limit of I-I-II correlators was computed by many methods \cite{Gromov:2011jh,Kostov:2012jr,Jiang:2016ulr,Kazama:2016cfl} and it is given by
\begin{multline}
    \log \left(\frac{C^{\bullet\bullet\bullet}}{C^{\circ\circ\circ}}\right) = \texttt{pols} + \oint_{\mathcal{C}_{1} \cup \mathcal{C}_{2}} \frac{dz}{2\pi} \textrm{Li}_2 \left( e^{i p_1 +i p_2 + i \ell_3 /2z } \right) + \oint_{\mathcal{C}_{3}} \frac{dz}{2\pi} \textrm{Li}_2 \left( e^{i p_3+ i (\ell_2 - \ell_1 ) /2z } \right) + \\
    - \frac{1}{2}\sum_{j=1}^{3} \oint_{\mathcal{C}_{j}} \frac{dz}{2\pi} \textrm{Li}_2 \left( e^{2i p_j} \right).
   \label{eq:claslim}
\end{multline}
where $C^{\circ\circ\circ}$ is the structure constant of three BPS operators with the same lengths as the excited operators and \texttt{pols} is just the terms containing all the polarizations which is fixed by SU(2) invariance. Such configuration is considerably simpler than I-I-I, since each sector has at most two interacting operators. However using a coherent-state approach, in \cite{Kazama:2016cfl} it was found the classical limit of the pure correlator:
\begin{multline}
    \log \left(\frac{C^{\bullet\bullet\bullet}}{C^{\circ\circ\circ}}\right) = \texttt{pols} + \frac{1}{2 }\sum_{\{i,j,k\} =  \text{cperm}\{1,2,3\}} \oint_{\mathcal{C}_i \cup \mathcal{C}_j} \frac{dz}{2\pi}\ \text{Li}_2\left(e^{ip_i + ip_j - ip_k}\right) + \\
    - \frac{1}{2}\sum_{j=1}^{3} \oint_{\mathcal{C}_{j}} \frac{dz}{2\pi} \textrm{Li}_2 \left( e^{2i p_j} \right).
   \label{eq:claslim2}
\end{multline}
Where cperm$\{1,2,3\}$ is all the cyclic permutations of $\{1,2,3\}$. Clearly the structure of pure and mixed correlators are similar, however a derivation directly from the microscopic expressions for the former was lacking. In Section \ref{sec:pure-clas-lim} we present such derivation.

In a similar vein we can also consider the classical limit in the SL(2) sector and try carrying the same analysis. Namely, we consider large spin $J_i \rightarrow \infty$ and large length $\ell\rightarrow \infty$, which implies big roots ($u \sim \ell$) \cite{Kazakov:2004nh}. Which is similar to the SU(2) case described earlier, however with some notable differences. The non-compact case can be seen as the analytic continuation in spin of the compact one. This simple modification make the Bethe roots real, \textit{id est}, the cuts are real. At strong coupling these large spin states corresponds to folded string solutions (or GKP strings) and in this limit their structure function is known \cite{Kazama:2011cp,Kazama:2012is}. 

Note that this is not the most general case for the classical limit in the SL(2) sector. One could consider finite length operators and asymptotically large spin since the chain is non-compact. Our expressions for the classical hexagon only work in the regime where both are large, thus we stick with this case at the moment. Although we do not consider it here,  the limit with finite length becomes important in the context of the WL/OPE duality, which we discuss in Sec. \ref{sec:sec5}. Nonetheless, a result in the SL(2) sector that is in our classical limit regime is \cite{Georgiou:2011qk}. Here it was computed the classical limit of SL(2) structure constants with two non-BPS operators consisting of lightcone derivatives and a BPS one. This means that in this case the polarizations are fixed. With this review and setup done, then let us discuss the new hexagon representations in the next section.

\section{A new representation for the hexagons}
\label{sec:new-hex-rep}

In this section we derive new expressions for the hexagon in both the compact and non-compact sectors. These new representations and their properties play a role in the classical limit in Section \ref{sec:sec3}.

\subsection{SU(2) Hexagons}
\label{sec:su2-newrep}

Here we will prove that the hexagon formulation and the double spin-chain approach yields the same result for the structure constant. This will yield us a new formula for the hexagons. Inspired by \cite{Kostov:2012jr}, we study the series expansion in $a_{\ell}$ of $\mathcal{G}$ and extract the first term which has no $a_{\ell}$'s. Such condition imposes the following constraints
\begin{equation}
    \bar{\beta}_1 \cap \alpha_2 = \bar{\beta}_2 \cap \alpha_3 = \bar{\beta}_3 \cap \alpha_1 = \beta_1 \cap \bar{\alpha}_1 = \beta_2 \cap \bar{\alpha}_2 = \beta_3 \cap \bar{\alpha}_3 = \emptyset.
\end{equation}
These conditions fix the subsets $\beta_{j}$ and $\bar{\beta}_j$ to be
\begin{equation}
    \beta_1 = \alpha_2, \ \beta_2 = \alpha_3, \ \beta_3 = \alpha_1, \ \bar{\beta}_1 = \bar{\alpha}_1, \ \bar{\beta}_2 = \bar{\alpha}_2, \ \bar{\beta}_3 = \bar{\alpha}_{3}.
\end{equation} 
Therefore, plugging them back in \eqref{eq:fullsc}, we get
\begin{multline}
    \mathcal{H} (\mathbf{u}_1, \mathbf{u}_2,\mathbf{u}_3) = (-1)^{M_1+M_2+M_3} \sum_{\substack{\alpha_1 \cup \Bar{\alpha}_1 = \mathbf{u}_1 \\ \alpha_2 \cup \Bar{\alpha}_2 = \mathbf{u}_2 \\ \alpha_3 \cup \Bar{\alpha}_3= \mathbf{u}_3}} z_{1 2}^{M_2 - M_3 - |\alpha_2|+ |\alpha_1| } z_{2 3}^{M_{3} - M_1 - |\alpha_3|+ |\alpha_2| } \times  \\ \times z_{3 1}^{M_1-M_2 - |\alpha_1|+ |\alpha_3| } f(\Bar{\alpha}_1,\alpha_1)f(\Bar{\alpha}_2,\alpha_2)f(\Bar{\alpha}_3,\alpha_3) f(\alpha_2,\bar{\alpha}_1)f(\alpha_3,\bar{\alpha}_2)f(\alpha_1,\bar{\alpha}_3).
    \label{G0}
\end{multline}
Note that the term above is very straightforward: it contains weights associated with each operator and interactions between `neighbor' partitions. The function $ \mathcal{H} (\mathbf{u}_1, \mathbf{u}_2,\mathbf{u}_3)$ will allow us to reorganize the terms in $\mathcal{G}$.

We can continue our procedure and analyze the term in $\mathcal{G}$ proportional to 
\begin{align}
a_{\ell_{31}}(\delta_1) a_{\ell_{12}}(\delta_2) a_{\ell_{23}}(\delta_3),
\end{align} 
where $\delta_i \cup \bar{\delta}_i = \mathbf{u}_i$. The partitions $\beta_i$ need to satisfy $\bar{\beta}_i \cap \alpha_j = \bar{\gamma}_j$ and $\beta_i \cap \bar{\alpha}_i = \gamma_i$ with $\gamma_i \cup \bar{\gamma}_i = \delta_i$, which fix them to
\begin{align}
    \beta_i = \alpha_j/\bar{\gamma}_j\cup\gamma_i \text{   and   } \bar{\beta}_i = \bar{\alpha}_i/\gamma_i \cup \bar{\gamma}_j .
\end{align}
Plugging this back in \eqref{eq:fullsc} yields us our first result\footnote{We defer the details of this computation to Appendix \ref{AppC}.}:
\begin{multline}
\mathcal{G}(\mathbf{u}_1,\mathbf{u}_2,\mathbf{u}_3) = \sum_{\substack{\delta_1 \cup \Bar{\delta}_1 = \mathbf{u}_1 \\ \delta_2 \cup \Bar{\delta}_2 = \mathbf{u}_2 \\ \delta_3 \cup \Bar{\delta}_3= \mathbf{u}_3}} (-1)^{|\delta_1|+|\delta_2|+|\delta_3|}  f(\bar{\delta}_1,\delta_1) f(\bar{\delta}_2,\delta_2) f(\bar{\delta}_3,\delta_3)   \\ \times  a_{\ell_{31}}(\delta_1) a_{\ell_{12}}(\delta_2) a_{\ell_{23}}(\delta_3) \mathcal{H}(\delta_1,\delta_3,\delta_2)\mathcal{H}(\bar{\delta}_1,\bar{\delta}_2,\bar{\delta}_3), \label{eq:G}
\end{multline}
which is the celebrated formula for the hexagonalization of structure constants from \cite{Basso:2015zoa}. Indeed, up to normalization factors, $\mathcal{H}(\delta_1,\delta_2,\delta_3)$ is a new representation for the hexagon for the SU(2) sector at tree level. Note that we proved the equivalence of hexagonalization and tailoring in full generality, thus extending the results of \cite{Basso:2015zoa}, which it was verified only for single excitation on each operator, and of \cite{Komatsu:2017buu}, which was done for a single non-BPS operator only. As we will explain in Section \ref{sec:sl2-newrep}, we will find a similar expression to \eqref{G0} for general polarized SL(2) correlators.

As in the SL(2) case, we can also derive recursion relations that the hexagon should satisfy. To derive these, we note that \eqref{G0} has a simple analytic structure: it is a rational function of the rapidities with poles for coincident roots. We see that there are no poles coming from two roots $x$ and $y$ from the same operator, since the only contributions are $(x \in \alpha_i, y \in \bar{\alpha}_i)$ and $(x \in \bar{\alpha}_i, y \in \alpha_i)$, which cancel each other. For roots from different operators, say $x \in \mathbf{u}_1$ and $y \in \mathbf{u}_2$, the only contribution to the residue $\text{Res}_{y \rightarrow x}\ \mathcal{H}$ comes from $x \in \bar{\alpha}_1$ and $y \in \alpha_2$. We can compute these residues and find the following recursion relations:
\begin{align}
    &\text{Res}_{y \rightarrow x} \ \mathcal{H}(\mathbf{u}_1 \cup \{x\},\mathbf{u}_2 \cup \{y\},\mathbf{u}_3) = i \ f(x,\mathbf{u}_1) f(\mathbf{u}_2,x) \mathcal{H}(\mathbf{u}_1,\mathbf{u}_2,\mathbf{u}_3), \label{eq:rec-rel1} \\
    &\text{Res}_{y \rightarrow x} \ \mathcal{H}(\mathbf{u}_1,\mathbf{u}_2 \cup \{x\},\mathbf{u}_3\cup \{y\}) = i \ f(x,\mathbf{u}_2) f(\mathbf{u}_3,x) \mathcal{H}(\mathbf{u}_1,\mathbf{u}_2,\mathbf{u}_3), \label{eq:rec-rel2}\\
    &\text{Res}_{y \rightarrow x} \ \mathcal{H}(\mathbf{u}_1 \cup \{y\},\mathbf{u}_2,\mathbf{u}_3 \cup \{x\}) = i \ f(x,\mathbf{u}_3) f(\mathbf{u}_1,x) \mathcal{H}(\mathbf{u}_1,\mathbf{u}_2,\mathbf{u}_3).\label{eq:rec-rel3}
\end{align}
These are the same recursion relations as in the SL(2) case up to the absence of polarizations and the change in sign, the latter due to the sign change at the pole in $f(u,v)$ from the SL(2) to SU(2) sector. Finally, if we take roots to infinity in \eqref{G0}, we can also show the remaining recursion relations
\begin{align}
    \text{lim}_{x \rightarrow \infty} \mathcal{H}(\mathbf{u}_1 \cup \{x\},\mathbf{u}_2,\mathbf{u}_3) = \mathcal{H}(\mathbf{u}_1,\mathbf{u}_2,\mathbf{u}_3), \label{lim1}\\
    \text{lim}_{x \rightarrow \infty} \mathcal{H}(\mathbf{u}_1,\mathbf{u}_2\cup \{x\},\mathbf{u}_3) = \mathcal{H}(\mathbf{u}_1,\mathbf{u}_2,\mathbf{u}_3), \label{lim2}\\
    \text{lim}_{x \rightarrow \infty} \mathcal{H}(\mathbf{u}_1 ,\mathbf{u}_2,\mathbf{u}_3\cup \{x\}) = \mathcal{H}(\mathbf{u}_1,\mathbf{u}_2,\mathbf{u}_3). 
    \label{lim3}
\end{align}
It is interesting to note that we could had work in the opposite order, i.e., by deriving first the recursion relations for the hexagon partition function in the SU(2) sector and then check that \eqref{G0} is indeed the solution of these. These relations will play a central role in deriving the classical limit of the hexagons in Section \ref{sec:sec3}. 

\subsection{SL(2) Hexagons}
\label{sec:sl2-newrep}

We saw that the SL(2) and SU(2) hexagons satisfy similar recursion relations and also that if we find a solution of them, this solution is unique. Based on this, we propose the following hexagon expression for the SL(2) case
\begin{multline}
    \label{eq:hexagonassum}
     \mathcal{H}(\mathbf{u}_1,\mathbf{u}_2,\mathbf{u}_3) = \sum_{\substack{\alpha_1 \cup \Bar{\alpha}_1 = \mathbf{u}_1 \\ \alpha_2 \cup \Bar{\alpha}_2 = \mathbf{u}_2 \\ \alpha_3 \cup \Bar{\alpha}_3= \mathbf{u}_3}} h_1^{|\alpha_1|} (1-h_1)^{|\bar{\alpha}_1|} h_2^{|\alpha_2|} (1-h_2)^{|\bar{\alpha}_2|} h_3^{|\alpha_3|} (1-h_3)^{|\bar{\alpha}_3|} \\ \times f(\Bar{\alpha}_1,\alpha_1)  f(\Bar{\alpha}_2,\alpha_2)f(\Bar{\alpha}_3,\alpha_3) f(\alpha_2,\bar{\alpha}_1)f(\alpha_3,\bar{\alpha}_2)f(\alpha_1,\bar{\alpha}_3),
\end{multline}
where $h_i$ depends on the polarizations in the following way
\begin{equation}
    h_2(1-h_1) = t_{12}, \ \ \ h_3(1-h_2) = t_{23}, \ \ \ \textrm{and} \ \ \  h_1(1-h_3) = t_{31}.
\end{equation}
With this it is simple to check that it satisfy all recursion relations. Then we have found a closed expression for the hexagon partition function. We remember that in \cite{Bercini:2022gvs} the hexagon partition function was a complicated sum over intermediate states in a Kagome lattice and remarkably \eqref{eq:hexagonassum} reproduces it exactly. It would be good in the future to derive this from the partition function approach. We comment more about it in Section \ref{sec:sec5}.

Just for completeness, now that we found the solution we can test it for some specific cases.  First, there are the very simple choices $(t_{12},t_{23},t_{31}) = (0,0,0)$ and $(t_{12},t_{23},t_{31}) = (1,0,0)$ related to the right- and left-sector of the I-I-II correlators in the SU(2) sector, respectively. For these polarizations the sum over partitions \eqref{eq:hexagonassum} truncate and we find
\begin{align}
    \mathcal{H}(\mathbf{u}_1,\mathbf{u}_2,\mathbf{u}_3) = 1 \ \ \ \textrm{and} \ \ \ \mathcal{H}(\mathbf{u}_1,\mathbf{u}_2,\mathbf{u}_3) = f(\mathbf{u}_2,\mathbf{u}_1),
\end{align}
Which matches the results in \cite{Bercini:2022gvs}. The latter, also known as the Abelian configuration, can be evaluated at finite coupling and, furthermore, its three-point function has a Pfaffian representation. For the $(t_{12},t_{23},t_{31}) = (0,0,0)$ case we can even find a closed form for the structure constant using the $\mathcal{A}$-functional (see Appendix \ref{AppB}), given by
\begin{equation}
    \label{eq:factorized}
    \mathcal{G} = \mathcal{A}_{\mathbf{u}_1}^{+}[a_{\ell_{31}}] \mathcal{A}_{\mathbf{u}_2}^{+}[a_{\ell_{12}}] \mathcal{A}_{\mathbf{u}_3}^{+}[a_{\ell_{23}}].
\end{equation}
Note that it is completely decoupled since the interactions between roots of distinct operators are given by the hexagons and these are trivial here.

Up to now, the previous discussed cases had simple analogs in SU(2). A slightly more general case we can consider is
$(t_{12},t_{23},t_{31}) = (t,0,0)$, where $0<t<1$. This choice translates to $(h_1,h_2,h_3) = (0,t,0)$ and then we have, using \eqref{eq:hexagonassum}, that
\begin{equation}
    \mathcal{H}(\mathbf{u}_1,\mathbf{u}_2,\mathbf{u}_3) = \sum_{\alpha_2 \cup \bar{\alpha}_2 = \mathbf{u}_2} t^{|\alpha_2|} (1-t)^{|\bar{\alpha}_2|} f(\bar{\alpha}_2,\alpha_2) f(\alpha_2,\mathbf{u}_1).
\end{equation}
Note that $\mathbf{u}_3$ decouples entirely and the $t=1$ and $t=0$ cases easily reduce to what we found before. Using again the $\mathcal{A}$-functional we can write this as a simple determinant
\begin{equation}
    \mathcal{H}(\mathbf{u}_1,\mathbf{u}_2,\mathbf{u}_3) = (1-t)^{|\mathbf{u}_2|} \mathcal{A}_{\mathbf{u}_2}^{+}\left[ \frac{t}{t-1} f (z,\mathbf{u}_1) \right].
\end{equation}
Therefore the correlator becomes simply
\begin{multline}
    \label{eq:spec-corr}
    \mathcal{G} = (1-t)^{M_2} \mathcal{A}_{\mathbf{u}_3}^{+}[a_{\ell_{23}}] \sum_{\substack{\alpha_1 \cup \Bar{\alpha}_1 = \mathbf{u}_1 \\ \alpha_2 \cup \Bar{\alpha}_2 = \mathbf{u}_2}} (-1)^{|\alpha_1|+|\alpha_2|} a_{\ell_{31}} (\alpha_1) a_{\ell_{12}} (\alpha_2) f(\Bar{\alpha}_1,\alpha_1) f(\Bar{\alpha}_2,\alpha_2) \\
    \mathcal{A}_{\alpha_2}^{+}\left[ \frac{t}{t-1} f (z,\alpha_1) \right] \mathcal{A}_{\Bar{\alpha}_2}^{+}\left[ \frac{t}{t-1} f (z,\Bar{\alpha}_1) \right].
\end{multline}
Which reduces in the $t\rightarrow 0$ limit to the previous factorized expression. Note that although we could not work out the sum over partitions in this case, the dependence on the polarizations is explicit unlike in the hexagon partition function. 

Note that we could write equations analogous to \eqref{scdef} and \eqref{Ddef} for the SL(2) three-point functions. In the compact sector, these equations come from a spin-chain framework, which is particularly useful for the computation of SU(2) three-point functions in the classical limit \cite{Kazama:2016cfl}. The idea is that, upon this limit, the Heisenberg spin-chain reduces to a Landau-Lifschitz sigma model and the correlators can be evaluated by using saddle-point and classical integrability tools. It is therefore a very interesting question if these methods can be generalized to general three-point functions in the SL(2) sector. We further discuss these in the concluding remarks.

\section{Classical hexagons}
\label{sec:sec3}

Here we will describe how to compute the classical limit of pure correlators in the SU(2) sector and of some fixed polarizations in the SL(2) case.

\subsection{Type I-I-I correlators}
\label{sec:pure-clas-lim}

Going back to the hexagon representation of $\mathcal{G}$ in \eqref{eq:G}, we see that the classical limit of all the quantities in the sum over partitions are known with the exception of the hexagon. To compute its classical limit we could work directly with representation \eqref{G0}, however we will consider here another approach. Consider two excited operators only, then the recursion relations \eqref{eq:rec-rel1} - \eqref{eq:rec-rel3} and \eqref{lim1} - \eqref{lim2} are solved by $\mathcal{H}(\mathbf{u}_1,\mathbf{u}_2) = f(\mathbf{u}_2,\mathbf{u}_1)$. Plugging this result in \eqref{eq:G}, we find the left side of I-I-II structure constants. Therefore a natural guess to solve the recurrence relations for three excited operators and type I-I-I correlators is
\begin{align}
    \mathcal{H}_{\text{try}}(\mathbf{u}_1,\mathbf{u}_2,\mathbf{u}_3) = f(\mathbf{u}_2,\mathbf{u}_1) f(\mathbf{u}_3,\mathbf{u}_2) f(\mathbf{u}_1,\mathbf{u}_3).
    \label{eq:Gtry}
\end{align}
This has the poles in the correct places and possess the appropriate asymptotic behavior. However, it does not satisfy the relations. Indeed for the first one we have:
\begin{align}
    \text{Res}_{y \rightarrow x}\ \mathcal{H}_{\text{try}}(\mathbf{u}_1 \cup \{x\},\mathbf{u}_2\cup \{y\},\mathbf{u}_3) = i  f(x,\mathbf{u}_1) f(\mathbf{u}_2,x) f(\mathbf{u}_3,x) f(x,\mathbf{u}_3)\mathcal{H}_{\text{try}}(\mathbf{u}_1,\mathbf{u}_2,\mathbf{u}_3).
\end{align}
Nevertheless, in the classical limit, where the roots are large and the sets of roots are well separated, we can use the following property
\begin{equation}
    f(\mathbf{u}_j,z) f(z,\mathbf{u}_j) \approx 1,
\end{equation}
with $z$ being a root in $\mathbf{u}_k$ with $k\neq j$. That is for \eqref{eq:Gtry} to be a solution of the recursion relations we must have the cuts to be macroscopically distant, e.g. the distance between roots $\mathbf{u}_1$ and $\mathbf{u}_3$ is of order $\mathcal{O}(\ell)$. Therefore, \textit{in the classical limit regime}, $\mathcal{H}$ is given by \eqref{eq:Gtry} for three excited operators. This derivation sounds a bit unnatural because we have mixed the microscopic analytic structure (recursion relations) of $\mathcal{H}$ with a macroscopic limit. It may be a good idea to study the classical limit from a more macroscopic point of view.

Given all the ingredients we now move to the computation of the full three-point function. Plugging our classical limit solution for $\mathcal{H}$ in \eqref{eq:G}, we have
\begin{align}
    \frac{\mathcal{G}(\mathbf{u}_1,\mathbf{u}_2,\mathbf{u}_3)}{\mathcal{H}(\mathbf{u}_1,\mathbf{u}_2,\mathbf{u}_3)} = \prod_{\{i,j,k\} = \text{cperm}\{1,2,3\}} \left(\sum_{\alpha_i \cup \bar{\alpha}_i= \mathbf{u}_i} (-1)^{|\alpha_i|} a_{\ell_{k i}}(\alpha_i) \frac{f(\bar{\alpha}_i,\alpha_i)}{f(\alpha_i,\mathbf{u}_k) f(\mathbf{u}_j,\alpha_i)}\right).
    \label{eq:disc-sum}
\end{align}
where cperm$\{1,2,3\}$ is all the cyclic permutations of $\{1,2,3\}$. The partitions being summed are now untangled, and the problem factorizes into three sums. In Appendix \ref{App:contour-manipulations}, we transform each of the sums into a path-integral and compute their classical limit using the methods in \cite{Escobedo:2010xs}.
In the end we can write the structure constant as:
\begin{equation}
    \log\mathcal{G}(\mathbf{u}_1,\mathbf{u}_2,\mathbf{u}_3) =  \frac{1}{2 }\sum_{\{i,j,k\} =  \text{cperm}\{1,2,3\}} \oint_{\mathcal{C}_i \cup \mathcal{C}_j} \frac{dz}{2\pi}\ \text{Li}_2\left(e^{ip_i + ip_j - ip_k}\right),
    \label{eq:final-limit}
\end{equation}
This result exactly matches the correlator \eqref{eq:claslim2} found in \cite{Kazama:2016cfl}\footnote{After we include the contributions from polarizations and Gaudin norms, the latter given in Appendix \ref{AppA}.}. Then starting from microscopic configurations we computed the classical limit of type I-I-I correlators. The approach here is completely orthogonal to the one in \cite{Kazama:2016cfl}, nonetheless the final result is the same, which confirms its validity.

\subsection{Spinning correlators}

Now we focus on the classical limit in the SL(2) sector. As aforementioned we are considering both spin $J$ and lengths $\ell$ large here. Unfortunately, we were only able to compute the classical limit of the hexagon for some special choices of the polarizations. For the choices of polarization $(t_{12},t_{23},t_{31}) = (0,0,0)$ and $(t_{12},t_{23},t_{31}) = (1,0,0)$ related to the mixed correlator the classical limit is the same as in the SU(2) case. A more complicated example we could consider is the analogous to the type I-I-I correlators, in which $(t_{12},t_{23},t_{31}) = (1,1,1)$. As before, we have
\begin{align}
    \mathcal{H}(\mathbf{u}_1,\mathbf{u}_2,\mathbf{u}_3) \approx f(\mathbf{u}_2,\mathbf{u}_1) f(\mathbf{u}_3,\mathbf{u}_2) f(\mathbf{u}_1,\mathbf{u}_3)
\end{align}
This is due to the fact that the recursion relations for this choice of polarizations is the same as in the pure correlators up to a sign related to the change in $f(u,v)$.

The more complicated case where we can analyze the classical limit is the one with polarizations $(t_{12},t_{23},t_{31}) = (t,0,0)$ and $0<t<1$, which was discussed before. Since there is no interaction between different partitions, the classical limit of the hexagon for these polarizations can be evaluated by path integral \cite{Escobedo:2011xw} or other methods \cite{Kostov:2012jr}. We find then in the classical limit
\begin{equation}
    \label{eq:hex-spec-class}
    \log\mathcal{H}(\delta_1,\delta_2,\delta_3) = \oint_{\mathcal{C}_2} \frac{du}{2 \pi}\ \text{Li}_2\left(\frac{t}{t-1} \ e^{i G_{\delta_1} (u) - i G_{\delta_2}(u)}\right),
\end{equation}
where $G_{\delta_j} (u)$ is the following resolvent defined by:
\begin{equation}
    G_{\delta_j} (u) = \int_{\mathcal{C}_j} \frac{dz}{2\pi}\ \frac{\rho_{\delta_j}(z)}{z-u}. 
\end{equation}
With this formula for the classical limit of the hexagon we can take it and insert into \eqref{eq:Gspinning} and compute the classical limit by the saddle-point method. Unfortunately we were not able to find the densities by solving the saddle-point equations due to their complexity. Indeed we have the following system of coupled integral equations for the densities:
\begin{multline}
    \log \left( \frac{\sinh(\pi(\rho_1 (v)- \rho_{\alpha_1} (v)))}{\sinh(\pi \rho_{\alpha_1} (v))} \right) + i \slashed{q}_1 (v) + \\
    \int_{\mathcal{C}_2} \frac{dx}{2\pi i}\ \frac{1}{v-x} \log\left( \frac{t-1 - t e^{iG_{\delta_1}(x) - iG_{\delta_2}(x)}}{t-1 - t e^{iG_{1}(x)-iG_{\delta_1}(x) - iG_{2}(x)+ iG_{\delta_2}(x)}} \right) + \\
    \int_{\mathcal{C}_2} \frac{dx}{2\pi i}\ \frac{1}{v-x} \log\left( \frac{t-1 - t e^{iG_{1}(x)-iG_{\delta_1}(x) + iG_{2}(x) - iG_{\delta_2}(x)-i\ell_{2}/x}}{t-1 - t e^{iG_{\delta_1}(x) + iG_{\delta_2}(x)-i\ell_{2}/x}} \right) = 0,
\end{multline}
\begin{multline}
    \log \left( \frac{\sinh(\pi(\rho_2 (v)- \rho_{\alpha_2} (v)))}{\sinh(\pi \rho_{\alpha_2} (v))} \right) + i \slashed{q}_2 (v) + \\
    \int_{\mathcal{C}_2} \frac{dx}{2\pi i}\ \frac{1}{v-x} \log\left( \frac{t-1 - t e^{iG_{1}(x)-iG_{\delta_1}(x) -iG_{2}(x)+ iG_{\delta_2}(x)}}{t-1 - t e^{iG_{\delta_1}(x) - iG_{\delta_2}(x)}} \right) + \\ 
    \int_{\mathcal{C}_2} \frac{dx}{2\pi i}\ \frac{1}{v-x} \log\left( \frac{t-1 - t e^{iG_{1}(x)-iG_{\delta_1}(x) +iG_{2}(x)-iG_{\delta_2}(x)-i\ell_{2}/x}}{t-1 - t e^{iG_{\delta_1}(x) + iG_{\delta_2}(x)-i\ell_{2}/x}} \right) = 0.
\end{multline}
Where $q_1$ and $q_2$ are the quasimomenta
\begin{equation}
    q_1 (u) = \pi +\frac{\ell_{31}}{u} + \int_{\mathcal{C}_1} \frac{dz}{2\pi}\ \frac{\rho_{1}(z)}{z-u} \ \ \ \textrm{and} \ \ \  q_2 (u) = \pi +\frac{\ell_{12}}{u} + \int_{\mathcal{C}_2} \frac{dz}{2\pi}\ \frac{\rho_{2}(z)}{z-u}.
\end{equation}
If we consider the limit $t\rightarrow 0$ we simply end with a system of two decoupled equations whose solution yields the classical limit of the factorized expression \eqref{eq:factorized}. Here we already see some of the complicated structures that arise in the classical limit of the SL(2) sector and have no parallel in the SU(2). Therefore it is highly desirable to have a new way to compute this classical limit which bypasses the need of solving the saddle-point equations. We discuss possibilities in the next section.

\section{Discussion and future directions}
\label{sec:sec5}

The hexagonalization formalism to compute structure constants in $\mathcal{N}=4$ SYM was an important achievement, however it is full of technical details that limit its applicability. Even if we consider the regime of large R-charge operators and bridges lengths, where the hexagon acquires its simplest form, we are still in practice limited to the study of operators with few excitations. This is due to the sheer complexity of the tensor contractions necessary to compute the hexagon form factors. The partition function interpretation for the hexagons introduced in \cite{Bercini:2022gvs} yielded us recursion relation that allowed for their complete determination at tree level, at least for spinning operators. 

Based on this work, here we wrote a general solution \eqref{eq:hexagonassum} of these recursion relations for the hexagon in the SL(2) sector. Now for the SU(2) sector we established the equivalence between tailoring and hexagonalization in full generality. As a byproduct of this we found recursion relations that resembles the ones in the non-compact sector and a new representation \eqref{G0} for the hexagon in the compact sector. These new hexagon expressions share a remarkable similar structure and hints at an unified description for all sectors. Also it reduces the complicated sum over configurations of the hexagon partition function in a Kagome lattice to a sum over partitions. 

Using these recursion relations we were also able to compute the classical limit of correlation functions in both SU(2) and SL(2) sectors. For the former we remember that in \cite{Kazama:2016cfl}, predictions were made for the classical limit of type I-I-I three-point functions
at weak coupling. However, it remained an open problem to reproduce the expressions found there from the microscopic sum over partitions \eqref{scdef}. Here we computed this classical limit with the aid of the aforementioned recursion relations and we confirm the predicted results. For the latter we used the same methods to compute the classical limit of structure constants, however we were able to do it only for some fixed polarizations that resembles correlators in the compact sector. For a slightly more general case we start to see some complicated structures that may appear. However not everything is settled and below we show some research prospects connected with this work which we think are relevant and could be fruitful for the future.

\subsection*{WL/OPE duality}

As suggested in \cite{Alday:2010ku}, the large spin limit of structure constants involving three spinning operators should be connected with null hexagonal Wilson loops (WL); a precise map for the kinematics of this duality was worked out in \cite{Bercini:2021jti}. This relation is the WL/OPE duality. It is one piece in a big web of dualities present in $\mathcal{N} = 4$ SYM, which involves polygonal Wilson loops, scattering amplitudes and three-point functions of spinning operators. All these quantities have different integrability descriptions: the polygonal Wilson loops and scattering amplitudes can be computed from the operator product expansion (OPE) of integrable pentagons, while three-point functions (and therefore polygonal Wilson loops) can be computed from the integrable hexagons. These describe the interactions of open and closed strings, respectively. To understand all these dualities from an integrability point of view and relate hexagons to pentagons is a very exciting challenge.

Our analysis may capture some of this large spin dynamics. Concretely, the polarizations considered in this work should be mapped to the polygonal Wilson loop with cross-ratios $(U_1,U_2,U_3) = (1,1,1)$, for $(t_{12}, t_{23}, t_{31}) = (1,1,1)$; and to $(U_1,U_2,U_3) = (0,0,0)$, for all other cases. In the context of Wilson loops, the case where the cross-ratios are null is singular; however, the former case where they are all equal to 1 corresponds to the null hexagonal Wilson loop which is regular  and can be embedded in AdS$_3$ \cite{Alday:2009dv}. However the duality in \cite{Bercini:2021jti} was established for the OPE of the smallest single-trace operators, therefore we are not in the limit of large R-charge that we explored in this work. To further apply our methods we need to see this duality in the limit of large R-charge or either modify the classical limit methods to account for large spin only. Also the lighlike Wilson loops at tree level are purely given by kinematics and therefore trivial, so we should also extend our methods to at least 1-loop for non-trivial results.

\subsection*{Classical and quantum integrability}

In our paper, we studied the microscopic description of three-point functions at weak coupling in the SL(2) and SU(2) sectors. In particular, we found that they can be computed from equations \eqref{eq:Gspinning} and \eqref{eq:G}, respectively. From them we calculated the classical limit of structure constants. However we can study these correlators directly in the macroscopic regime. Indeed, for SU(2), in the classical limit, there exists a macroscopic description in which this problem is mapped to a Landau-Lifschitz model and can be solved by using classical integrability tools at weak and strong coupling \cite{Kazama:2016cfl}. In this language, these weak coupling correlators have the same structure as the strong coupling ones, which describe the scattering of three classical closed strings rotating in  S$^3$.

In the future, we hope to explore the possibility of mapping the computation of correlators in the full SL(2) sector at weak coupling to a classically integrable sigma model as in the SU(2) case. In that language, these objects should be closely related to their strong coupling counterparts, which describe the three-point function of closed strings rotating in AdS$_5$. With this we hope to bypass all the trouble of finding the saddle-point equations and solving them in the classical limit.

Having in mind the WL/OPE duality, we expect this problem to be very similar to the computation of Wilson loops living in AdS$_5$, in which the result can be expressed as the free energy of a thermodynamics Bethe ansatz system \cite{Alday:2009dv}. Note that for three-point functions at strong coupling and with polarizations on a plane, this was already computed in \cite{Kazama:2011cp,Kazama:2012is}, which would serve as a test for any proposal. A good starting point for achieving this goal may be to use the spin-vertex to compute the three-point functions and use their monodromy relations to compute their classical limit \cite{Jiang:2014cya}.

\subsection*{Hexagon partition function and  all-loop generalizations}

As said before we found a new representation \eqref{eq:hexagonassum} for the spinning hexagons of \cite{Bercini:2022gvs}. We found a similar expression \eqref{G0} for the SU(2) sector also, although there is no counterpart as a partition function. So a good direction would be to analyze the hexagon partition function in all generality and look for the recursion relations in all sectors, at least at tree level initially. Maybe a generalization of \eqref{G0} and \eqref{eq:hexagonassum} is feasible. As a non-trivial exercise, it would also be important to prove the equivalence between the partition function representation and our expressions for the hexagons, which could maybe hint at some new non-trivial relations of the partition function.

We could also look beyond tree level. As explained in \cite{Bercini:2022gvs}, at finite coupling the hexagon partition function is not a rational function and it possess branch cuts introduced by the Zhukovsky variables. Therefore the recursion relations are not enough to fix it at finite coupling. Also we have extra poles at finite coupling lying in the other sheets that are related to bound states, which are not considered in the recursion relations. A possible direction to analyze our expressions at 1-loop is to use the $\theta$-morphism, which was used in the tailoring formalism to describe structure constants at higher loops \cite{Gromov:2012uv,Gromov:2012vu}.
We conclude by noting that an all-loop generalization of \eqref{eq:hexagonassum} may not be at all out-of-reach. Indeed, we found a simple generalization of it given
\begin{multline}
    \mathcal{H}(\delta_1,\delta_2,\delta_3) = \hat{h}(\delta_2^{2 \gamma},\delta_1) \hat{h}(\delta_1^{2 \gamma},\delta_3) \hat{h}(\delta_3^{2 \gamma},\delta_2)  \sum_{\alpha_i \cup \Bar{\alpha}_i = \delta_i} \prod_{k=1}^{3} h_k^{|\alpha_k|} (1-h_k)^{|\bar{\alpha}_k|}  \\ 
    \prod_{\{i,j,k\} =  \text{cperm}\{1,2,3\}}\frac{1}{\hat{h}(\Bar{\alpha}_i,\alpha_i)\hat{h}(\Bar{\alpha}_i^{2 \gamma},\alpha_i)\hat{h}(\alpha_j,\bar{\alpha}_i)\hat{h}(\alpha_j^{2 \gamma},\bar{\alpha}_i)},
\end{multline}
with $\hat{h}(u,v)$ and $\hat{h}(u^{2 \gamma},v)$ being the hexagon scalar factor and its analytic continuation given in \cite{Basso:2015zoa} (stripped of BES dressing factor), respectively. Also at tree level $f(u,v)=1/\hat{h}(u,v)$. This expression passes various tests, it reduces to \eqref{eq:hexagonassum} at tree level, it satisfies the decoupling relations at \textit{finite} coupling and it reduces to the abelian case in the appropriate limit. Although it passes these non-trivial tests, it is not a polynomial in $t_{ij}$ for generic polarizations beyond tree level and therefore it must be corrected. We hope in the future to be able to make progress in these matters.

\acknowledgments

We thank Pedro Vieira, for suggesting this topic and inspiration. We thank Alexandre Homrich, Shota Komatsu, Ivan Kostov, and Didina Serban for illuminating discussions. This work was supported by CAPES, FAPESP grant 2019/12167-3, and ICTP-SAIFR FAPESP grant 2016/01343-7.


\appendix

\section{Gaudin norm of Bethe states} 
\label{AppA}

The norm of a Bethe state is called the Gaudin norm. It can be computed directly from the algebraic Bethe ansatz \cite{Korepin:1982gg}. Given a set of roots $\mathbf{u}$ that describes a certain Bethe state, its norm is
\begin{equation}
    \mathcal{N} (\mathbf{u}) = \sqrt{\prod_{x\in \mathbf{u}} \left( x-i/2 \right)^{\ell} \prod_{y\in \mathbf{u}}  \left( y+ i/2 \right)^{\ell}}  \ \mathcal{B}(\mathbf{u}),
\end{equation}
where $\ell$ is the length of the spin-chain and
\begin{align}
    & \mathcal{B}(\mathbf{u})= \sqrt{\prod_{n\neq m} \left( \frac{u_n - u_m + i}{u_n - u_m}\right) \det \mathcal{M}}\ ,
    \\
    & \mathcal{M}_{nm} = \frac{2}{(u_n - u_m)^2 + 1} + \delta_{nm} \biggl( \frac{\ell}{u_n^2 + 1/4} -\sum_{i=1}^M \frac{2}{(u_n - u_i)^2 + 1} \biggr).
\end{align}
The polynomial factors in the norm can be absorbed in the definition of $\mathcal{D}(\alpha_j, \bar{\alpha}_j)$ as we do in Appendix \ref{AppB}. So it remains to compute the classical limit of the determinant factor $\mathcal{B}(\mathbf{u})$. It was already found in \cite{Gromov:2011jh, Kostov:2012jr} and it is just   
\begin{equation}
   \log \mathcal{B}(\mathbf{u})= \frac{1}{2} \oint_{\mathcal{C}_{\mathbf{u}}} \frac{dz}{2\pi} \textrm{Li}_2 \left( e^{2i p} \right).
\end{equation}
This explain the presence of the single contour terms in equations \eqref{eq:claslim} and \eqref{eq:claslim2}. 


\section{Bethe equations and rewriting $\mathcal{D}(\alpha_j, \bar{\alpha}_j)$}
\label{AppB}

In this appendix we rewrite $\mathcal{D}(\alpha_j, \bar{\alpha}_j)$ using Bethe equations such that the final structure constant will then be independent of the polarizations $z_{ij}$ for off-shell roots. First we define the Baxter polynomials as usual and a different normalization for the contraction of Bethe states:
\begin{align}
    & Q^{\pm}_{\ell} (\mathbf{x}) = \prod_{u\in\mathbf{x}} (u\pm i/2)^{\ell} , \\
    & Z_p (\alpha_{i} \cup \Bar{\alpha}_{j} |\ell_{ij}) = Q^{+}_{\ell_{ij}} ( \Bar{\alpha}_{j} ) Q^{-}_{\ell_{ij}} (\alpha_{i}) \tilde{Z}_p (\alpha_{i} \cup \Bar{\alpha}_{j} |\ell_{ij}).
\end{align}
This rewriting of the scalar products suits our purposes since  it is $\tilde{Z}_p (\alpha_{i} \cup \Bar{\alpha}_{j} |\ell_{ij})$ that has a good classical limit and not the former definition. We can write $\mathcal{D}(\alpha_j, \bar{\alpha}_j)$ as
\begin{multline}
   \mathcal{D}(\alpha_j, \bar{\alpha}_j) =  (-1)^{|\alpha_{1}|+|\alpha_{2}|+|\alpha_{3}|} \prod\limits_{i=1}^{3} f(\alpha_{i} , \Bar{\alpha}_{i}) \prod\limits_{i=1}^{3} Q^{-}_{\ell_i} (\alpha_{i}) Q^{+}_{\ell_i} (\Bar{\alpha}_{i}) \tilde{Z}_p (\alpha_{1} \cup \Bar{\alpha}_{3} |\ell_{13}) \times \\   \tilde{Z}_p (\alpha_{2} \cup \Bar{\alpha}_{1} |\ell_{12})  \tilde{Z}_p (\alpha_{3} \cup \Bar{\alpha}_{2} |\ell_{23}).
\end{multline}
Now one uses the $\mathcal{A}$-functional defined in \cite{Kostov:2012jr,Escobedo:2010xs} by
\begin{equation}
    \mathcal{A}_{\mathbf{u}}^{\pm}[g] = \frac{\det_{ab} \left( u_a^{b-1} - g(u_a) (u_a \pm i)^{b-1} \right)}{\det_{ab} (u_a^{b-1})},
\end{equation}
which has a sum over partitions expression given by
\begin{equation}
    \mathcal{A}_{\mathbf{u}}^{\pm}[g] = \sum_{\alpha\cup\Bar{\alpha} = \mathbf{u}} (-1)^{|\alpha|} g(\alpha) f(\alpha,\Bar{\alpha}).
\end{equation}
Where $+$ is for the SU(2) $f(u,v)$ and $-$ for the SL(2) one. We note that the choice $g(z) = a_{\ell}(z)$ yields  the determinant part of the pDWPF and it is the closed form expression for the scalar product of a Bethe state and a descendant of the vacuum. Let $\mathcal{A}_{\mathbf{u}}^{+}[a_{\ell}] = \mathcal{A} (\mathbf{u} |\ell)$, then
\begin{multline}
   \mathcal{D}(\alpha_j, \bar{\alpha}_j) =  (-1)^{|\alpha_{1}|+|\alpha_{2}|+|\alpha_{3}|} \prod\limits_{i=1}^{3} f(\alpha_{i} , \Bar{\alpha}_{i}) \frac{\prod\limits_{i=1}^{3} Q^{-}_{\ell_i} (\alpha_{i}) Q^{+}_{\ell_i} (\Bar{\alpha}_{i})}{a_{\ell_{13}}(\Bar{\alpha}_{3}) a_{\ell_{12}}(\Bar{\alpha}_{1}) a_{\ell_{23}}(\Bar{\alpha}_{2})} \mathcal{A} (\alpha_{1} \cup \Bar{\alpha}_{3} |\ell_{13})  \times \\   
   \mathcal{A} (\alpha_{2} \cup \Bar{\alpha}_{1} |\ell_{12}) \mathcal{A} (\alpha_{3} \cup \Bar{\alpha}_{2} |\ell_{23}).
\end{multline}
It remains to divide the structure constant by the norms of the operators. As said in Appendix \ref{AppA} the determinant part of the norm has a well defined classical limit, then we define $\tilde{\mathcal{D}}(\alpha_j, \bar{\alpha}_j)$ as the previous summand but divided by the polynomial part of the norms. Therefore
\begin{multline}
   \tilde{\mathcal{D}}(\alpha_j, \bar{\alpha}_j) =  \frac{(-1)^{|\alpha_{1}|+|\alpha_{2}|+|\alpha_{3}|}}{{a_{\ell_{13}}(\Bar{\alpha}_{3}) a_{\ell_{12}}(\Bar{\alpha}_{1}) a_{\ell_{23}}(\Bar{\alpha}_{2})}} \prod\limits_{i=1}^{3} f(\alpha_{i} , \Bar{\alpha}_{i}) \prod\limits_{i=1}^{3} \sqrt{\frac{a_{\ell_i} (\Bar{\alpha}_{_i})}{a_{\ell_i} (\alpha_{_i})}} \ \mathcal{A} (\alpha_{1} \cup \Bar{\alpha}_{3} |\ell_{13}) \times \\ 
   \mathcal{A} (\alpha_{2} \cup \Bar{\alpha}_{1} |\ell_{12}) \mathcal{A} (\alpha_{3} \cup \Bar{\alpha}_{2} |\ell_{23}).
\end{multline}
Since we are dealing with physical operators, the total momentum of each one vanishes. Then
\begin{equation}
    a_{\ell_j} (\mathbf{u}_{j}) = 1.
\end{equation}
We can take this identity and redefine the variables in the sum and obtain the new summand
\begin{multline}
   \tilde{\mathcal{D}}(\alpha_j, \bar{\alpha}_j) =  (-1)^{|\alpha_{1}|+|\alpha_{2}|+|\alpha_{3}|}  \prod\limits_{i=1}^{3} f(\alpha_{i},\Bar{\alpha}_{i}) a_{\ell_{23}}(\Bar{\alpha}_{3}) a_{\ell_{13}}(\Bar{\alpha}_{1}) a_{\ell_{12}}(\Bar{\alpha}_{2}) \ \mathcal{A} (\alpha_{1} \cup \Bar{\alpha}_{3} |\ell_{13}) \times  \\
   \mathcal{A} (\alpha_{2} \cup \Bar{\alpha}_{1} |\ell_{12}) \mathcal{A} (\alpha_{3} \cup \Bar{\alpha}_{2} |\ell_{23}).
\end{multline}
This is now ready to be manipulated and obtain equation (\ref{defDinv}).

The full structure constant $\mathcal{G}$ is $z_{ij}$ independent if the roots satisfy the Bethe equations. Using the sum over partitions expression for the $\mathcal{A}$-functional, the full summand is just
\begin{multline}
    \tilde{\mathcal{D}} = (-1)^{|\bar{\alpha}_1| + |\bar{\alpha}_2| + |\bar{\alpha}_3|} \sum_{\substack{\beta_1 \cup \bar{\beta}_1 = \bar{\alpha}_1 \cup \alpha_2 \\ \beta_2 \cup \bar{\beta}_2 = \bar{\alpha}_2 \cup \alpha_3 \\ \beta_3 \cup \bar{\beta}_3 = \bar{\alpha}_3 \cup \alpha_1}}(-1)^{|\beta_1| + |\beta_2| + |\beta_3|} \frac{a_{\ell_{12}}(\alpha_2) a_{\ell_{23}}(\alpha_3) a_{\ell_{31}}(\alpha_1)}{a_{\ell_{12}}(\beta_1) a_{\ell_{23}}(\beta_2) a_{\ell_{31}}(\beta_3)} \\ \times f(\bar{\alpha}_1,\alpha_1)f(\bar{\alpha}_2,\alpha_2) f(\bar{\alpha}_3,\alpha_3) f(\beta_1,\bar{\beta}_1) f(\beta_2,\bar{\beta}_2) f(\beta_3,\bar{\beta}_3).
\end{multline}
Note that:
\begin{equation}
   \frac{a_{\ell_{12}}(\alpha_2) a_{\ell_{23}}(\alpha_3) a_{\ell_{31}}(\alpha_1)}{a_{\ell_{12}}(\beta_1) a_{\ell_{23}}(\beta_2) a_{\ell_{31}}(\beta_3)}  = \frac{a_{\ell_{12}}(\bar{\beta}_1 \cap \alpha_2) a_{\ell_{23}}(\bar{\beta}_2 \cap \alpha_3) a_{\ell_{31}}(\bar{\beta}_3 \cap \alpha_1)}{a_{\ell_{12}}(\beta_1 \cap \bar{\alpha}_1)  a_{\ell_{23}}(\beta_2 \cap \bar{\alpha}_2)a_{\ell_{31}}(\beta_3 \cap \bar{\alpha}_3)}.
\end{equation}
Clearly $\beta_j \cap \bar{\alpha}_j \subset \mathbf{u}_{j}$ then one can use the Bethe equations to invert $a_{\ell}(x)$ in the denominator. In our notation the Bethe equations are simply
\begin{equation}
    a_{\ell_{j}}(x) = S(\{x\},\mathbf{u}_{j}).
\end{equation}
In the end we obtain the final form of the summand in equation (\ref{defDinv}) 
\begin{multline}
    \tilde{\mathcal{D}}(\alpha_j, \bar{\alpha}_j)  = (-1)^{|\bar{\alpha}_1| + |\bar{\alpha}_2| + |\bar{\alpha}_3|} \sum_{\substack{\beta_1 \cup \bar{\beta}_1 = \bar{\alpha}_1 \cup \alpha_2 \\ \beta_2 \cup \bar{\beta}_2 = \bar{\alpha}_2 \cup \alpha_3 \\ \beta_3 \cup \bar{\beta}_3 = \bar{\alpha}_3 \cup \alpha_1}}(-1)^{|\beta_1| + |\beta_2| + |\beta_3|} a_{\ell_{12}}(\bar{\beta}_1 \cap \alpha_2) a_{\ell_{23}}(\bar{\beta}_2 \cap \alpha_3)\\
    a_{\ell_{31}}(\bar{\beta}_3 \cap \alpha_1) a_{\ell_{31}}(\beta_1 \cap \bar{\alpha}_1) a_{\ell_{12}}(\beta_2 \cap \bar{\alpha}_2) a_{\ell_{23}}(\beta_3 \cap \bar{\alpha}_3)\\ S\left(\mathbf{u}_{1},\beta_1 \cap\bar{\alpha}_1\right) S\left(\mathbf{u}_{2},\beta_2 \cap\bar{\alpha}_2\right)
    S\left(\mathbf{u}_{3},\beta_3 \cap\bar{\alpha}_3\right)\\ f(\bar{\alpha}_1,\alpha_1)f(\bar{\alpha}_2,\alpha_2) f(\bar{\alpha}_3,\alpha_3) f(\beta_1,\bar{\beta}_1) f(\beta_2,\bar{\beta}_2) f(\beta_3,\bar{\beta}_3).
\end{multline} 
Similar manipulations were made in \cite{Basso:2015zoa} to check that $\mathcal{G}$ is independent of the polarizations. However they only tested it for one excitation on each operator. Here we work for a general number of them and establish the above relation.


\section{Simplifying hexagon recursion relations}
\label{App:redefinitions}

Recently in \cite{Bercini:2022gvs} some recursion relations were found for spinning hexagons. In this appendix we will rewrite them in a more appropriate form for us. Let $\mathcal{Z}(\mathbf{u},\mathbf{v},\mathbf{w})$ denote the hexagon partition function with three excited operators with $J_i$ excitations each one. The recursion relations found \cite{Bercini:2022gvs} take the following form:
\begin{multline}
    \text{Res}_{v_j \rightarrow u_i} \mathcal{Z}(\mathbf{u},\mathbf{v},\mathbf{w}) = i \langle L_1, R_2 \rangle \langle L_2, R_1 \rangle \prod_{i^\prime = 1}^{i-1} \frac{f(u_i,u_{i^\prime})}{f(u_{i^\prime},u_i)} \prod_{j^\prime = j+1}^{J_2} \frac{f(v_{j^\prime},u_i)}{f(u_i,v_{j^\prime})} \times \\ 
    \times \mathcal{Z}(\mathbf{u}/\{u_i\},\mathbf{v}/\{v_j\},\mathbf{w}),
\end{multline}
\begin{align}
    \lim_{u_i \rightarrow \infty} \mathcal{Z}(\mathbf{u},\mathbf{v},\mathbf{w}) = (-1)^{J_1+J_2+J_3+1} \langle L_1, R_1 \rangle  \times \mathcal{Z}(\mathbf{u}/\{u_i\},\mathbf{v},\mathbf{w}).
\end{align}
It turns out we can greatly simplify these recursion relations by redefining the hexagon partition functions as
\begin{align}
    \mathcal{H}(\mathbf{u},\mathbf{v},\mathbf{w}) = \mathcal{Z}(\mathbf{u},\mathbf{v},\mathbf{w}) \times
    \frac{f_<(\mathbf{u}_1) f_<(\mathbf{u}_2) f_<(\mathbf{u}_3)}{V_1^{J_1} V_2^{J_2} V_3^{J_3}} \times (-1)^{\frac{(J_1+J_2+J_3)(J_1+J_2+J_3-1)}{2}},
\end{align}
where
\begin{equation}
    f_{<}(\mathbf{u}) = \prod_{i<j} \left(\frac{u_i-u_j-i}{u_i - u_j} \right).
\end{equation}
The first recursion relation for $\mathcal{H}(\mathbf{u},\mathbf{v},\mathbf{w})$ then become
\begin{multline}
  \frac{1}{f_<(\mathbf{u}_1) f_<(\mathbf{u}_2/\{v_j\}) f_<(\mathbf{u}_3)} \prod_{j^\prime = 1}^{j-1} \frac{1}{f(v_{j^\prime},u_i)} \prod_{j^\prime = j+1}^{J_2} \frac{1}{f(u_i,v_{j^\prime})} \times \text{Res}_{v_j \rightarrow u_i} \mathcal{H}(\mathbf{u},\mathbf{v},\mathbf{w}) = \\ - i t_{12} \prod_{i^\prime = 1}^{i-1} \frac{f(u_i,u_{i^\prime})}{f(u_{i^\prime},u_i)} \prod_{j^\prime = j+1}^{J_2} \frac{f(v_{j^\prime},u_i)}{f(u_i,v_{j^\prime})} \times \frac{\mathcal{H}(\mathbf{u}/\{u_i\},\mathbf{v}/\{v_j\},\mathbf{w})}{f_<(\mathbf{u}_1/\{u_i\}) f_<(\mathbf{u}_2/\{v_j\}) f_<(\mathbf{u}_3)},
\end{multline}
which, after simplifications, it becomes \eqref{eq:eq1-rr}:
\begin{align}
   \text{Res}_{v_j \rightarrow u_i}\mathcal{H}(\mathbf{u},\mathbf{v},\mathbf{w}) = 
   - i t_{12} f(u_i,\mathbf{u}_1/\{u_i\}) f(\mathbf{u}_2/\{v_j\},u_i) \mathcal{H}(\mathbf{u}/\{u_i\},\mathbf{v}/\{v_j\},\mathbf{w}).
\end{align}
And the second recursion relation becomes
\begin{align}
    \lim_{u_i \rightarrow \infty} \mathcal{H}(\mathbf{u},\mathbf{v},\mathbf{w}) = \mathcal{H}(\mathbf{u}/\{u_i\},\mathbf{v},\mathbf{w}),
\end{align}
Which is exactly \eqref{eq:dec1} in Section \ref{sec:spinning-hex}. The remaining recursion relations are obtained with similar manipulations.


\section{Redefining the sum over partitions}
\label{AppC}

In this appendix we rewrite the sum over partitions \eqref{eq:fullsc} to reproduce the hexagon formula \eqref{eq:G}. As said in the Section \ref{sec:su2-newrep} we define the partitions $\{\gamma_j , \bar{\gamma}_j \}$ such that $\bar{\beta}_i \cap \alpha_j = \bar{\gamma}_j$ and $\beta_i \cap \bar{\alpha}_i = \gamma_i$ where $\gamma_i \cup \bar{\gamma}_i = \delta_i$ and we partition each operator's roots as $\delta_i \cup \bar{\delta}_i = \mathbf{u}_i$. This is defined such that the all the terms containing $a_{\ell}$'s in \eqref{defDinv} are joined. Then the partitions $\{\beta_j , \bar{\beta}_j \}$ are given by
\begin{align}
    \beta_i = \alpha_j/\bar{\gamma}_j\cup\gamma_i \text{   and   } \bar{\beta}_i = \bar{\alpha}_i/\gamma_i \cup \bar{\gamma}_j.
\end{align}
Applying this at $\tilde{\mathcal{D}}$ results in 
\begin{multline}
    \tilde{\mathcal{D}} = (-1)^{|\bar{\delta}_1| + |\bar{\delta}_2| + |\bar{\delta}_3|} \times a_{\ell_{31}}(\delta_1) a_{\ell_{12}}(\delta_2) a_{\ell_{23}}(\delta_3)  \\ \times \frac{f(\bar{\gamma}_1,\gamma_1)}{f(\gamma_1,\bar{\gamma}_1)} \frac{f(\bar{\gamma}_2,\gamma_2)}{f(\gamma_2,\bar{\gamma}_2)} \frac{f(\bar{\gamma}_3,\gamma_3)}{f(\gamma_3,\bar{\gamma}_3)}
    \frac{f(\bar{\delta}_1,\gamma_1)}{f(\gamma_1,\bar{\delta}_1)} \frac{f(\bar{\delta}_2,\gamma_2)}{f(\gamma_2,\bar{\delta}_2)} \frac{f(\bar{\delta}_3,\gamma_3)}{f(\gamma_3,\bar{\delta}_3)}
    \\ \times
    f(\bar{\alpha}_1/\gamma_1,\alpha_1) f(\gamma_1,\alpha_1) f(\bar{\alpha}_2/\gamma_2,\alpha_2) f(\gamma_2,\alpha_2) f(\bar{\alpha}_3/\gamma_3,\alpha_3) f(\gamma_3,\alpha_3)
    \\ \times f(\alpha_2/\bar{\gamma}_2,\bar{\alpha}_1/\gamma_1) f(\gamma_1,\bar{\alpha}_1/\gamma_1) f(\alpha_2/\bar{\gamma}_2,\bar{\gamma}_2) f(\gamma_1,\bar{\gamma}_2)
    \\ \times f(\alpha_3/\bar{\gamma}_3,\bar{\alpha}_2/\gamma_2) f(\gamma_2,\bar{\alpha}_2/\gamma_2) f(\alpha_3/\bar{\gamma}_3,\bar{\gamma}_3) f(\gamma_2,\bar{\gamma}_3)
    \\ \times f(\alpha_1/\bar{\gamma}_1,\bar{\alpha}_3/\gamma_3) f(\gamma_3,\bar{\alpha}_3/\gamma_3) f(\alpha_1/\bar{\gamma}_1,\bar{\gamma}_1) f(\gamma_3,\bar{\gamma}_1).
\end{multline}
After some algebraic manipulations one can see that is useful to define partitions for the set $\bar{\delta}_{i}=\omega_i \cup \bar{\omega}_i$ such that $\alpha_{i} = \bar{\omega}_i \cup \bar{\gamma}_{i}$ and $\bar{\alpha}_{i} = \omega_i \cup \gamma_{i}$. This yields us the following simple expression
\begin{multline}
    \tilde{\mathcal{D}} = (-1)^{|\bar{\delta}_1| + |\bar{\delta}_2| + |\bar{\delta}_3|}  \times a_{\ell_{31}}(\delta_1) a_{\ell_{12}}(\delta_2) a_{\ell_{23}}(\delta_3) \times f(\bar{\delta}_1,\delta_1) f(\bar{\delta}_2,\delta_2) f(\bar{\delta}_3,\delta_3) \\ \times
    f(\omega_1,\bar{\omega}_1) f(\bar{\omega}_1,\omega_3) f(\omega_3,\bar{\omega}_3) f(\bar{\omega}_3,\omega_2) f(\omega_2,\bar{\omega}_2) f(\bar{\omega}_2,\omega_1) \\ \times   f(\bar{\gamma}_1,\gamma_1) f(\gamma_1,\bar{\gamma}_2)  f(\bar{\gamma}_2,\gamma_2) f(\gamma_2,\bar{\gamma}_3) f(\bar{\gamma}_3,\gamma_3) f(\gamma_3,\bar{\gamma}_1).
\end{multline}
Which has the same structure as the one in $\mathcal{H}$. Now the only remaining factors in the sum over partitions to deal with are the polarizations. For these we can do the following split:
\begin{align}
    & z_{1 2}^{M_{12} - |\alpha_2|- |\bar{\alpha}_1| } = z_{12}^{|\bar{\delta}_1|+|\bar{\delta}_2|-|\bar{\delta}_3| - |\bar{\omega}_2| - |\omega_1|} z_{12}^{|\delta_1|+|\delta_2|-|\delta_3| - |\bar{\gamma}_2| - |\gamma_1| }, \\ 
    & z_{23}^{M_{23} - |\alpha_3|- |\bar{\alpha}_2| } = z_{23}^{|\bar{\delta}_2|+|\bar{\delta}_3|-|\bar{\delta}_1| - |\bar{\omega}_3| - |\omega_2|} z_{12}^{|\delta_2|+|\delta_3|-|\delta_1| - |\bar{\gamma}_3| - |\gamma_2|}, \\
    & z_{31}^{M_{31} - |\alpha_1|- |\bar{\alpha}_3| } = z_{31}^{|\bar{\delta}_3|+|\bar{\delta}_1|-|\bar{\delta}_2| - |\bar{\omega}_1| - |\omega_3|} z_{31}^{|\delta_3|+|\delta_1|-|\delta_2| - |\bar{\gamma}_1| - |\gamma_3|}.
\end{align}
Thus each one can be absorbed in the sum over partitions $\gamma_i \cup \bar{\gamma}_i = \delta_i$ and $\omega_i \cup \bar{\omega}_i = \bar{\delta}_i$ to obtain exactly:
\begin{multline}
\mathcal{G}(\mathbf{u}_1,\mathbf{u}_2,\mathbf{u}_3) = \sum_{\substack{\delta_1 \cup \Bar{\delta}_1 = \mathbf{u}_1 \\ \delta_2 \cup \Bar{\delta}_2 = \mathbf{u}_2 \\ \delta_3 \cup \Bar{\delta}_3= \mathbf{u}_3}} (-1)^{|\delta_1|+|\delta_2|+|\delta_3|}  f(\bar{\delta}_1,\delta_1) f(\bar{\delta}_2,\delta_2) f(\bar{\delta}_3,\delta_3)   \\ \times  a_{\ell_{31}}(\delta_1) a_{\ell_{12}}(\delta_2) a_{\ell_{23}}(\delta_3) \mathcal{H}(\delta_1,\delta_3,\delta_2)\mathcal{H}(\bar{\delta}_1,\bar{\delta}_2,\bar{\delta}_3).
\end{multline}
With $\mathcal{H}$ given by \eqref{G0}. This establishes the equality between double spin-chain formalism and hexagonalization.


\section{Contour manipulations in the classical limit}
\label{App:contour-manipulations}

In this appendix we show how to find the final form of the classical limit of type I-I-I correlators \eqref{eq:final-limit}. The first step is compute the classical limit by saddle-point of path integrals as in \cite{Gromov:2011jh}. We can substitute each sum in \eqref{eq:disc-sum} by path integrals of the form
\begin{align}
    \sum_{\alpha_i \cup \bar{\alpha}_i= \mathbf{u}_i} (-1)^{|\alpha_i|} a_{\ell_{k i}}(\alpha_i) \frac{f(\bar{\alpha}_i,\alpha_i)}{f(\alpha_i,\mathbf{u}_k) f(\mathbf{u}_j,\alpha_i)} \rightarrow \int [D\rho_{\alpha_i}] \exp\left(\int_{\mathcal{C}_i} du\ \mathcal{F}[\rho_{\alpha_i}] +   i \rho_{\alpha_i} \slashed{q}_i \right).
\end{align}
Where $\mathcal{F}[\rho_{\alpha_i}]$ is the stochastic anomaly term, given by 
\begin{align}
    \mathcal{F}[\rho_{\alpha_i}] = F(\rho_{i}) - F(\rho_{\alpha_i}) - F(\rho_{\bar{\alpha}_i})  \ \ \ \textrm{with} \ \ \ F(\rho) = \int_{0}^{\rho} dx\ \log\sinh(\pi x),
\end{align}
and $i \rho_{\alpha_i} \slashed{q}_i$ is the classical limit of the summand. Here, $q_i$ is a function of the quasimomenta given by
\begin{align}
    q_i = \pi + p_j - p_k - p_i,
\end{align}
and $\slashed{q}_i$ denotes the average of $q_i$ evaluated above and below the cut $\mathcal{C}_i$. 

Then one does these path integrals by saddle-point and finds
\begin{align}
     \log \frac{\mathcal{G}(\mathbf{u}_1,\mathbf{u}_2,\mathbf{u}_3)}{\mathcal{H}(\mathbf{u}_1,\mathbf{u}_2,\mathbf{u}_3)} = - \sum_{\{i,j,k\} =  \text{cperm}\{1,2,3\}} \oint_{\mathcal{C}_i} \frac{du}{2\pi} \ \text{Li}_2\left(e^{-ip_k - ip_i + ip_j}\right).
     \label{almostShota}
\end{align}
To find the same as in \cite{Kazama:2016cfl} we have to join the contours $\mathcal{C}_i$. To do this we use Bethe equations in the classical limit and contour manipulations. The Bethe equations in the classical limit are given by \cite{Kazakov:2004qf}
\begin{equation}
    p_{j}(u+i0) + p_{j}(u-i0) = 2\pi n^{(k)}_j \ \ \ \textrm{for} \ \ \  u\in\mathcal{C}^{(k)}_{j}.
\end{equation}
Where $\cup_{k} \mathcal{C}^{(k)}_{j} = \mathcal{C}_{j}$ and $n^{(k)}_j \in \mathbb{Z}$. The Bethe equations tell us that $p(u)$ is defined in a two sheet Riemann surface. Then when one crosses a rapidity cut we simply do the changes $p(u)\rightarrow -p(u)$ and $\mathcal{C}_{j}\rightarrow - \mathcal{C}_{j}$.

Consider the first term in \eqref{almostShota}. We start by splitting it as
\begin{equation}
    \oint_{\mathcal{C}_1} \frac{du}{2\pi} \ \text{Li}_2\left(e^{-ip_3 - ip_1 + ip_2}\right) = -\frac{1}{2} \oint_{\mathcal{C}_1} \frac{du}{2\pi} \ \text{Li}_2\left(e^{ip_1 + ip_2 - ip_3}\right) 
    + \frac{1}{2} \oint_{\mathcal{C}_1} \frac{du}{2\pi} \ \text{Li}_2\left(e^{-ip_3 - ip_1 + ip_2}\right),
\end{equation}
where in the first term we used the Bethe equations to go to the second sheet, thus changing the sign of $p_1$. We consider now the following dilogarithm identity 
\begin{equation}
    \textrm{Li}_2 (x) = - \textrm{Li}_2 (-1/x) - \frac{\pi^2}{6} - \frac{\log^2 (1/x)}{2}. 
\end{equation}
Which can be used to invert $p_j$ in the second term thus yielding
\begin{multline}
    \oint_{\mathcal{C}_1} \frac{du}{2\pi} \ \text{Li}_2\left(e^{-ip_3 - ip_1 + ip_2}\right) = -\frac{1}{2} \oint_{\mathcal{C}_1} \frac{du}{2\pi} \ \text{Li}_2\left(e^{ip_1 + ip_2 - ip_3}\right) 
    - \frac{1}{2} \oint_{\mathcal{C}_1} \frac{du}{2\pi} \ \text{Li}_2\left(e^{ip_1 + ip_3 - ip_2}\right) \\
    - \frac{1}{4} \oint_{\mathcal{C}_1} \frac{du}{2\pi} \ \log^2\left(-e^{ip_3 + ip_1 - ip_2}\right).
\end{multline}
Similar manipulations can be done in the remaining terms in \eqref{almostShota} to obtain
\begin{equation}
    \log \frac{\mathcal{G}(\mathbf{u}_1,\mathbf{u}_2,\mathbf{u}_3)}{\mathcal{H}(\mathbf{u}_1,\mathbf{u}_2,\mathbf{u}_3)} = \frac{1}{2 }\sum_{\{i,j,k\} =  \text{cperm}\{1,2,3\}} \left(\oint_{\mathcal{C}_i \cup \mathcal{C}_j} \frac{dz}{2\pi}\ \text{Li}_2\left(e^{ip_i + ip_j - ip_k}\right)\right) + \texttt{rest}.
\end{equation}
Where \texttt{rest} is simply given by
\begin{equation}
    \texttt{rest} = \sum_{\{i,j,k\} =  \text{cperm}\{1,2,3\}} \frac{1}{4} \oint_{\mathcal{C}_i} \frac{du}{2\pi} \ \log^2\left(-e^{ip_i - ip_j + ip_k}\right).
\end{equation}
After some painstaking, but simple, manipulations one sees that $\log\mathcal{H}(\mathbf{u}_1,\mathbf{u}_2,\mathbf{u}_3)$ indeed cancels with \texttt{rest} up to phases that can be absorbed by redefining the operators. Therefore the classical limit \eqref{almostShota} is indeed equivalent to the classical limit \eqref{eq:final-limit} given in the main text.



\bibliographystyle{unsrt}
\bibliography{main}

\end{document}